\documentclass[graybox]{svmult}
\usepackage{type1cm}        
\usepackage{makeidx}         
\usepackage{graphicx}        
\usepackage{multicol}        
\usepackage[bottom]{footmisc}
\usepackage{newtxtext}       
\usepackage[varvw]{newtxmath}       

\makeindex             

\begin{document}

\title*{Thermodynamic Flux-Force Closure Relations for Systems out of the Onsager Region\\
\author{\hskip 4.5cm{\bf Giorgio SONNINO}}
\institute{{\bf Giorgio Sonnino}, {\it{Universit{\'e} Libre de Bruxelles (ULB), Department of Physics, Bvd du Triomphe, Campus de la Pliane, C.P. 231 - Build. NO - 1050 Brussels (Belgium)}}
\\
\email{giorgio.sonnino@ulb.be}
}
}
\maketitle

\hskip 1.9cm{\it{Manuscript Accepted for publication in Springer Nature}}
\vskip 1truecm
\abstract{
The first attempts to develop non-equilibrium thermodynamics theory occurred after the first observations of some coupled phenomena of thermal diffusion and thermoelectric. However, the big obstacle to overcome is that the number of unknowns is greater than the number of equations expressing the conservation laws. So, it is crucial to determine the \textit{closure relations} to make the problem solvable. The objective of this work is to determine the nonlinear flux-force relations for systems out of Onsager’s region that respect the existing thermodynamic theorems for systems far from equilibrium. To this aim, a thermodynamic theory for irreversible processes [referred to as the \textit{Thermodynamical Field Theory} (TFT)] has been developed. The TFT rests upon the concept of equivalence between thermodynamic systems. The equivalent character of two alternative descriptions of a thermodynamic system is ensured if, and only if, the two sets of thermodynamic forces are linked with each other by the so-called \textit{Thermodynamic Covariant Transformations} (TCT). The TCT are the most general thermodynamic force transformations which leave unaltered both the entropy production and the Glansdorff-Prigogine dissipative quantity. In this work, we describe the Lie group and the group representations associated to the TCT. The TCT leads to the so-called \textit{Thermodynamic Covariance Principle} (TCP): \textit{The nonlinear closure equations, i.e., the flux-force relations, must be covariant under} TCT. In this chapter, we provide the explicit form of the nonlinear PDEs, subjected to the appropriate boundary conditions, which have to be satisfied by transport coefficients when the skew-symmetric piece is absent. The solution of these equations allows to determine the flux-force closure relations for systems out of the Onsager region. Since the proposed PDEs are obtained without neglecting any term present in the balance equations (i.e., the mass, momentum, and energy balance equations), we propose them as a good candidate for describing transport in thermodynamic systems also in turbulent regime. As a special case, we derive the nonlinear PDEs for transport coefficients when the thermodynamic system is subjected to two thermodynamic forces. The obtained PDE is, in Thermodynamical Field Theory (TFT), analogous to Liouville's equation in Riemannian (or pseudo-Riemannian) geometry. A preliminary test is carried out by analysing a concrete example where Onsager's relations manifestly disagree with experience: losses in magnetically confined Tokamak-plasmas. More specifically, we compute the mass and energy losses in FTU (\textit{Frascati Tokamak Upgrade})-plasmas subjected to two thermodynamic forces. We show the good agreement between the theoretical (TFT) predictions and the experimental data. The aim is to apply our approach to the \textit{Divertor Tokamak Test facility} (DTT), to be built in Italy, and to ITER. Other applications of the TFT to the thermoelectric effects or to out-of-equilibrium chemical reactions can be found in the references cited at the end of the chapter.
} 

\section{Introduction}\label{I}
When there are more unknowns than equations expressing conservation laws additional \textit{closure laws} are needed to make the problem solvable. Generally, these additional closure relations are not derivable from one of the physical equations being solved. Several approaches to getting the closure relations are currently applied. Among them we cite the so called \textit{troncation schemes} and the \textit{Asymptotic schemes}. In \textit{truncation schemes}, higher order moments are arbitrarily assumed to vanish, or simply negligible with respect to the terms of lower moments. Truncation schemes can often provide quick insight into fluid systems, but always involve uncontrolled approximation. This method is often used in transport processes in Tokamak-plamas (see, for instance, the book \cite{balescu2}). The \textit{asymptotic schemes} are based on the rigorous exploitation of some small parameter. They have the advantage of being systematic, and providing some estimate of the error involved in the closure. However, as the title itself suggests, these methods are effective only when small parameters enter, by playing a crucial role, in the dynamic equations. These schemes are often used for solving numerically kinetic equations (ref., for instance, to the book \cite{carillo}). Another possibility is to obtain the closure relations by formulating a specific theory or \textit{ad hoc} models. The most important closure equations are the so-called \textit{transport equations} (or the \textit{flux-force relations}), relating the thermodynamic forces with the conjugate dissipative fluxes that produce them. The thermodynamic forces are related to the spatial inhomogeneity and (in general) they are expressed as gradients of the thermodynamic quantities. The study of these relations is the object of \textit{non-equilibrium thermodynamics}. Morita and Hiroike eased this task for a closure relation by providing the formally exact closure formula \cite{morita}
\begin{equation}\label{I1}
J_\nu(X)=\varpi_{\mu\nu}(X)X^\mu
\end{equation}
\noindent Here, $X^\mu$ and $J_\mu$ denote the thermodynamic forces and thermodynamic fluxes, respectively. Coefficients $\varpi_{\mu\nu}(X)$ are the \textit{transport coefficients}, where it is clearly highlighted that the transport coefficients may depend on the thermodynamic forces. We suppose that all quantities appearing in Eq.~(\ref{I1}) are dimensionless. Note that in this equation, as well as in the sequel, the Einstein summation convention on the repeated indices is understood. Matrix $\varpi_{\mu\nu}(X)$ can be decomposed into a sum of two matrices, one symmetric and the other skew symmetric, which we denote with $g_{\mu\nu}(X)$ and $f_{\mu\nu}(X)$, respectively. The second law of thermodynamics requires that $g_{\mu\nu}(X)$ is a positive-definite matrix. Note that, in general, the dimensionless entropy production, denoted by $\sigma$, with $\sigma=\varpi_{\mu\nu}(X) X^\mu X^\nu =g_{\mu\nu}(X)X^\mu X^\nu$, may not be a simply bilinear expression of the thermodynamic forces (since the transport coefficients may depend on the thermodynamic forces). For conciseness, in the sequel we drop the symbol $X$ in $g_{\mu\nu}$ as well as in the skew-symmetric piece of the transport coefficients $f_{\mu\nu}$ being implicitly understood that these matrices may depend on the thermodynamic forces. 

\noindent In previous works, a macroscopic \textit{Thermodynamic Field Theory} (TFT) for deriving the closure relations valid for thermodynamic systems out of Onsager's region has been proposed. More specifically, the aim of the TFT in \cite{sonnino}-\cite{sonnino1} is to determine the nonlinear flux-force relations which are valid for thermodynamic systems out of the thermodynamic linear region (commonly referred to as the \textit{Onsager region}) \cite{onsager1}, \cite{onsager2}. This task is accomplished by means of three hypotheses: two constraints 1. and 2., and one assumption 3. In order to establish the vocabulary and notations that shall be used in the sequel of this work, we briefly recall these hypotheses.

\begin{enumerate}
\item {\textit{The thermodynamic laws and the theorems demonstrated for systems far from equilibrium must be satisfied.}}
\item{\textit{The validity of the Thermodynamic Covariance Principle (TCP) must be ensured}. 

\noindent The TCP stems from the concept of equivalent systems from the thermodynamic point of view. Thermodynamic equivalence was originally introduced by Th. De Donder and I. Prigogine \cite{prigogine1}, \cite{prigogine2}, \cite{degroot}. However, the De Donder-Prigogine definition of thermodynamic equivalence, based only on the invariance of the entropy production, is not sufficient to guarantee the equivalence character between two sets of thermodynamic forces and conjugate thermodynamic fluxes. In addition, it is known that there exists a large class of flux-force transformations such that, even though they leave unaltered the expression of the entropy production, they may lead to certain paradoxes \cite{Verschaffelt}, \cite{Davies}. The equivalent character of two alternative descriptions of a thermodynamic system is ensured if, and only if, the two sets of thermodynamic forces are linked with each other by the so-called \textit{Thermodynamic Covariant Transformations} (TCT). The TCT are the most general thermodynamic force transformations which leave unaltered both the entropy production $\sigma$ and the Glansdorff-Prigogine dissipative quantity $P$ [for the definition of $P$, see the forthcoming Eq. (\ref{RDDP1})]. In this work, we also describe the Lie group and the group representations associated to the TCT. The TCT leads in a natural way to postulate the validity of the so-called \textit{Thermodynamic Covariance Principle} (TCP): \textit{The nonlinear closure equations, i.e., the flux-force relations must be covariant under the Thermodynamic Covariant Transformations} (TCT).}
\item{\textit{Close to the steady states, the nonlinear closure equations can be derived by the principle of least action.}}
\end{enumerate}
\noindent This theory, based on 1., 2., and 3. is referred to as the \textit{Thermodynamical Field Theory} (TFT). The three hypotheses 1., 2., and 3. allow determining the nonlinear TFT-Partial Differential Equations (PDEs) for transport coefficients $\varpi_{\mu\nu}$. In this chapter, we shall limit ourselves to the case in which the transport coefficients possess only the symmetric piece (i.e., $f_{\mu\nu} = 0$). We shall show the explicit form of the TFT-PDEs for $g_{\mu\nu}$.  Later, inspired by the theory of Jackiw and Teitelboim \cite{jackiw}-\cite{teitelboim}, we shall also derive the explicit form of the TFT-PDEs for $g_{\mu\nu}$ for the two-dimensional case, i.e. when the system is subjected to two independent thermodynamic forces.

\noindent The final part of the chapter is devoted to the application of the theory to some relevant examples of systems out of equilibrium. More precisely, we shall apply the derived TFT-PDEs to Tokamak-plasmas in collisional regime. This is a very interesting example of application since, in this case, the Onsager relations strongly disagree with experimental data. One of the main issues in Fusion Science is the computation of energy and mass losses in Tokamak-plasmas. It is well-known that there is a strong disagreement, of several orders of magnitude, especially for electron mass and energy losses, between the theoretical predictions of the Onsager theory (at the basis of the so-called \textit{neoclassical theory}) and experiments. This discrepancy is even more pronounced in case of magnetically confined plasmas in turbulent regime. The aim is to compute the electron heat loss in Tokamak-plasmas by considering the contribution of the non-linear terms in the flux-force relations derived by the TFT. In order to test the validity of our results, we have computed the electron heat loss  for FTU (Frascati Tokamak Upgrade)-plasmas in fully collisional transport regime. We have compared the theoretical profile obtained by the nonlinear theory satisfying the TCP with the experimental data for the FTU-plasmas (provided by the ENEA C.N.R. - EuroFusion in Frascati) and with the theoretical predictions of the linear theory (the Onsager theory). We found that there is a fairly good agreement between the TFT and experiments (in contrast with Onsager's theory). However, disagreements appear in the region where the dimensionless entropy production $\sigma$ is of order $1$. In particular, we found that the disagreement appears in the region of the tokamak where the plasma is in turbulent regime. Incidentally, this corresponds also to the region where $\sigma\sim 1$. Preliminary calculations and theoretical results in the region $\sigma\sim 1$ have also been performed. We are currently comparing the theoretical predictions of the TFT with the experimental data for FTU-plasma in turbulent regime. Other examples of application of the TFT to unimolecular triangular chemical reactions (i.e., three isomerisations take place) and to materials subjected to temperature and electric potential gradients, and to chemical reactions out of equilibrium and to Hall-effect can be found in refs~\cite{sonnino} and \cite{sonnigi4} and in \cite{sonnigi7}-\cite{sonnino3}, respectively.

\noindent The chapter is organised as follows. In Section~\ref{TFT} we recall the basic concepts of the Thermodynamical Field Theory (TFT). To this aim, we quickly introduce the definition of the \textit{space of the thermodynamic forces} and we describe the \textit{Thermodynamic Covariance Principle} (TCP) and the TCT-\textit{symmetry group}. Successively, in Section~\ref{nte} we derive the explicit form of the nonlinear TFT-PDEs for transport coefficients in absence of the skew-symmetric part (i.e., when $f_{\mu\nu}=0$). In Section~\ref{2dim} we derive the nonlinear PDEs for transport coefficients $g_{\mu\nu}$ when the system is subjected to two independent thermodynamic forces (i.e., when $n=2$). Section~\ref{LTE} provides the linearised TFT transport equations. The physical meaning of the gauge invariance in TFT is reported in Section~\ref{TFTGI}. The analytic solution of the 2-dimensional linearised homogeneous TFT-PDE is obtained in Section~\ref{linsol1}. The solution of the TFT-PDE for collisional FTU-plasmas, subjected to two independent thermodynamic forces, can be found in Section~\ref{solution}. In this Section we shall show the good agreement between the theoretical predictions and the experimental data. Finally, the main results are concluded in Section~\ref{conclusions}. Here we also specify the boundary conditions for turbulent Tokamak-plasmas. The determination of the boundary conditions, which have to be satisfied by the TFT-PDEs for a general system out of thermodynamic equilibrium, are obtained in the Appendix. In Appendix we can also find the analytic solutions of the linearised, 2-dimensional, inhomogeneous TFT-PDE.

\section{The Thermodynamical Field Theory (TFT)}\label{TFT} 
\subsection{The Space of the Thermodynamic Forces}
The first task is to define the space where we may perform calculations. To this aim, it is not enough to specify the nature of the axes, we must also determine two quantity: the {\it metric tensor} and the {\it affine connection} (denoted with symbol $\Gamma^\lambda_{\mu\nu}$). 

\noindent - The metric tensor is a central object in the theory; it describes the local geometry of space. The metric tensor is a symmetric tensor used to raise and lower the indicative tensors and generate the connections used to construct the PDEs and the curvature tensor of the space.

\noindent - The curvature of a space can be identified by taking a vector at some point and transporting it parallel along a curve in space-time. An affine connection is a rule that describes how to legitimately move a vector along a curve on the variety without changing its direction.

\begin{figure}[b]
\sidecaption
\includegraphics[scale=.60]{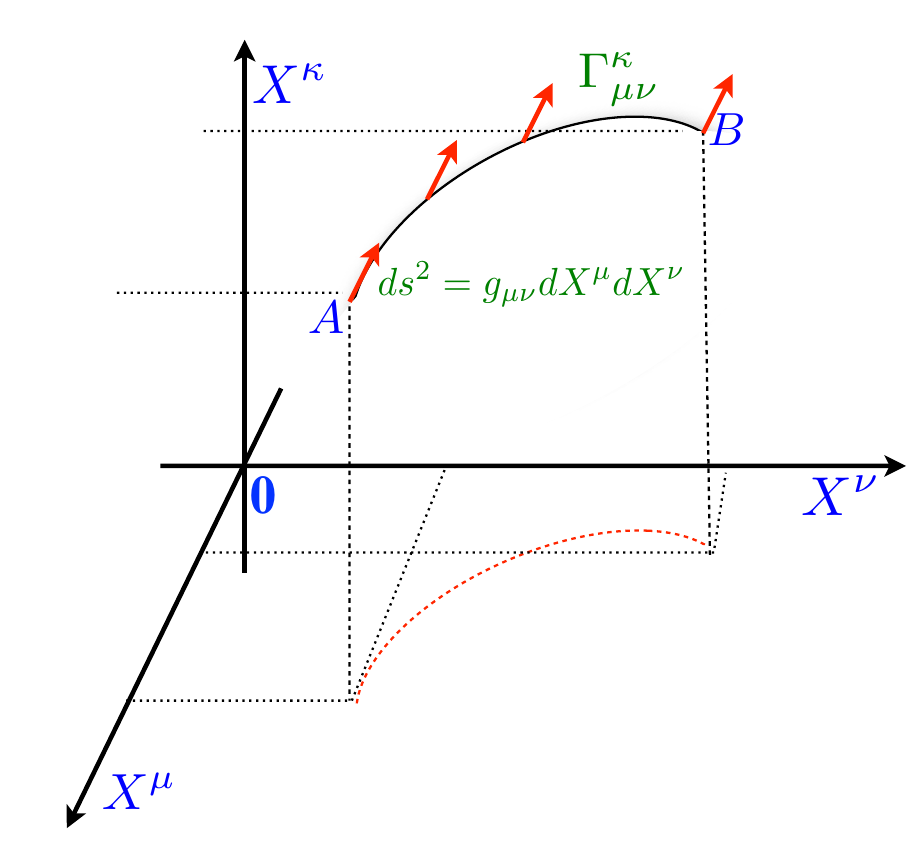}
\caption{{\bf The Thermodynamic Space}. The space is spanned by the thermodynamic forces, the metric tensor is identified with the symmetric piece of the transport coefficients, and the expression of the affine connection is determined by the {\it General Evolution Criterion}.}
\label{fig_Axes}
\end{figure}

\noindent The metric tensor and the affine connection are determined by physics, i.e., by ensuring the validity of the thermodynamic theorems valid for systems out of equilibrium (in accordance with the above-mentioned assumption 1. More precisely, we must take into account the validity of the \textit{second law of thermodynamics} and the \textit{General Evolution Criterion} (GEC) \cite{prigogine3}, \cite{prigogine4}. We adopt the following definitions \cite{sonnino}, \cite{sonnino1}: 
\begin{itemize}
\item{{\it The space of the thermodynamic forces} (or, simply, {\it the thermodynamic space}) \textit{is the space spanned by the thermodynamic forces}}. 
\end{itemize}
\begin{itemize}
\item{\textit{The metric tensor is identified with the symmetric piece} $g_{\mu\nu}$ \textit{of the transport coefficients}.}
\begin{itemize}
\item{Note that this definition takes into account the second law of thermodynamics as, for the second law of thermodynamics, the square (infinitesimal) distance $ds^2=d{\bf s}\cdot d{\bf s}$ is always a non negative quantity  - see Fig.~(\ref{fig_Axes}).}
\end{itemize}
\end{itemize}
\begin{itemize}
\item{\textit {The expression of the thermodynamic affine connection} ${\widetilde\Gamma}^\kappa_{\mu\nu}$ \textit{is determined by requiring the validity of the Glansdorff-Prigogine General Evolution Criterion}. In ref.~\cite{sonnino1} it is shown that, when $f_{\mu\nu}=0$, we get:}
\end{itemize}
\begin{align}\label{TFT1}
{\widetilde\Gamma}^\mu_{\alpha\beta}\equiv&
\begin{Bmatrix} 
\mu \\ \alpha\beta
\end{Bmatrix}
\!+\!\frac{1}{2\sigma}X^\mu X^\eta g_{\alpha\beta,\eta}&\nonumber \\ 
&-\frac{1}{2(n+1)\sigma}\Bigl(
\delta^\mu_\alpha X^\nu X^\eta g_{\beta\nu,\eta}+\delta^\mu_\beta X^\nu X^\eta g_{\alpha\nu,\eta}\Bigr)
\end{align}
\noindent where commas stand for partial derivatives with respect to the thermodynamic forces. $\delta_i^j$ denotes the Kronecker delta and
\begin{equation}
\begin{Bmatrix} 
\mu \\ \alpha\beta
\end{Bmatrix}=\frac{1}{2}g^{\mu\lambda}\Bigl(\frac{\partial g_{\lambda\alpha}}{\partial X^\beta}+\frac{\partial g_{\lambda\beta}}{\partial X^\alpha}-\frac{\partial g_{\alpha\beta}}{\partial X^\lambda}\Bigr)
\end{equation}
\noindent the \textit{Levi-Civita affine connection}.

\subsection{The De Donder-Prigogine Thermodynamic Invariance}\label{DDP}
Onsager's theory is based on three assumptions : i) The probability distribution function for the fluctuations of thermodynamic quantities (Temperature, pressure, degree of advancement of a chemical reaction etc.) is a Maxwellian, ii) Fluctuations decay according to a linear law, and iii) The detailed balance principle (or the microscopic reversibility) is satisfied. Onsager showed the equivalence of the assumptions i), ii), and iii) with the equations \cite{onsager1}, \cite{onsager2}
\begin{equation}\label{DDP1}
J_\nu=L_{\mu\nu}X^\mu\qquad {\rm with}\qquad \frac{\partial L_{\mu\nu}}{\partial X^\lambda} =0\qquad {\rm and}\qquad L_{\mu\nu}=L_{\nu\mu}
\end{equation}
\noindent where the coefficients of matrix $L_{\mu\nu}$ are the Onsager transport coefficients. $L_{\mu\nu}$ is a symmetric matrix and the elements are independent of the thermodynamic forces. The limit of validity of Eq.~({\ref{DDP1}) establishes the limit of validity of the \textit{Onsagr region}. Assumption {\bf iii)} allows deriving the \textit{reciprocity relations} $L_{\mu\nu} = L_{\nu\mu}$. The Onsager theory of fluctuations starts from the Einstein formula linking the probability of a fluctuation, $W$, with the entropy change, $\Delta S$, associated with the fluctuations from the state of equilibrium
\begin{equation}\label{DDP2}
W=W_0\exp [\Delta S/K_B]
\end{equation}
\noindent In Eq.~(\ref{DDP2}), $K_B$ is the Bolzmann constant and $W_0$ is a normalization constant which ensures that the sum of all probabilities equals to one \cite{onsager1}, \cite{onsager2}. Prigogine generalized Eq.~(\ref{DDP2}), which applies to adiabatic or isothermal transformations, by introducing the entropy production due to fluctuations. Denoting with $\xi_i$ ($i=1\cdots m$) the $m$ deviations of the thermodynamic quantities from their equilibrium value, Prigogine proposed that the probability distribution of finding a state in which the values $\xi_i$ lie between $\xi_i$ and $\xi_i + d\xi_i$ is given by \cite{prigogine2}
\begin{equation}\label{DDP3}
W=W_0\exp [\Delta_I S/K_B]\qquad \Delta_I S =\int_E^F d_Is\quad ;\quad \frac{d_Is}{dt}\equiv \int_\Omega \sigma d {\rm v}
\end{equation}
\noindent Here, $d{\rm v}$ is the spatial volume element of the system, and the integration is over the entire space $\Omega$ occupied by the system. $E$ and $F$ indicate the equilibrium state and the state to which a fluctuation has driven the system, respectively. Note that this \textit{probability distribution remains unaltered for flux‐force transformations leaving invariant the entropy production}. Concrete examples of chemical reactions, equivalent from the thermodynamic point of view, have also been analysed in literature. As an example, among these, we choose the simplest of all. Let us consider, for example, the following chemical system in which two isomerisations 

{\bf a)}: $A\rightarrow B$ and $B\rightarrow C$.

\noindent take place \cite{prigogine2}. Of course, from the macroscopic point of view, the chemical changes in {\bf a)} are equivalent to the two isomerisations

 {\bf b)}: $A\rightarrow C$ and $B\rightarrow C$.

\noindent It can be checked that, under a linear transformation of the thermodynamic forces (which in this case corresponds to a linear transformation of the chemical affinities) the entropy productions for the two chemical reactions {\bf a)} and {\bf b)}, are equal. Indeed, the corresponding affinities of the reactions {\bf a)} read: $A^1= \mu_A-\mu_B$ and $ A^2=\mu_B-\mu_C$, with $A^i$ and $\mu_i$ ($i=A,B,C$) denoting the \textit{chemical affinities} and the \textit{chemical potentials}, respectively. The change per unit time of the mole numbers is given by
\begin{equation}\label{DDP4}
\frac{dn_A}{dt}=- v_1\quad ;\quad \frac{dn_B}{dt}= v_1-v_2\quad ;\quad \frac{dn_C}{dt}= v_2
\end{equation}
\noindent with $v_i$ ($i=1,2$) denoting the chemical reaction rates. In this case the thermodynamic forces and the thermodynamic fluxes are the chemical affinities (over temperature) and the chemical reaction rates, respectively i.e., $X^\mu=A^\mu/T$ and $J_\mu =v_\mu$. Hence, the corresponding entropy production reads $d_I S/dt=(A^1/T)\nu_1+(A^2/T)\nu_2\geq 0$. The affinities corresponding to reactions {\bf b)} are related to the previous ones by
\begin{equation}\label{DDP5}
A^{'1}=\mu_A-\mu_C=A^1+A^2\quad ;\quad A^{'2}=\mu_B-\mu_C=A^2
\end{equation}
\noindent By taking into account that
\begin{equation}\label{DDP6}
\frac{dn_A}{dt}=- v{'}_1\quad ;\quad \frac{dn_B}{dt}= -v{'}_2\quad ;\quad \frac{dn_C}{dt}= v{'}_1+v{'}_2
\end{equation}
\noindent we get
\begin{equation}\label{DDP7}
v_1= v{'}_1\quad ;\quad v_2=v{'}_1+v{'}_2
\end{equation}
\noindent where the invariance of the entropy production is manifestly shown. Indeed,
\begin{equation}\label{DDP8}
\frac{d_IS}{dt}=(A^1/T)v_1+ (A^2/T)v_2=(A^{'1}/T)v{'}_1+ (A^{'2}/T)v{'}_2=\frac{d_IS{'}}{dt}
\end{equation}
\noindent or $J_\mu X^\mu = J_{\mu}^{'} X^{'\mu}$ (where, as usual, the Einstein summation convention on the repeated indexes is adopted). On the basis of the above observations, Th. De Donder and I. Prigogine formulated, for the first time, the concept of equivalent systems from the thermodynamical point of view. For Th. De Donder and I. Prigogine, thermodynamic systems are thermodynamically equivalent if, under flux-force transformation, the bilinear form of the entropy production remains unaltered i.e., $\sigma =\sigma {'}$ \cite{prigogine1}, \cite{prigogine2}, \cite{degroot}.
\subsection{Remarks on De Donder-Prigogine's Thermodynamic Invariance Formulation }\label{RDDP}
The Thermodynamic Invariance Principle formulated by De Donder‐Prigogine, based only on the invariance of the entropy production, is not sufficient to guarantee the equivalence character of the two descriptions $(J_\mu,X^\mu)$ and $(J_\mu^{'},X^{'\mu})$. Indeed, we can easily convince ourselves that there exists a large class of transformations such that, even though they leave unaltered the expression of the entropy production, they may lead to certain paradoxes to which J.E. Verschaffelt and R.O. Davies have called attention \cite{Verschaffelt}, \cite{Davies}. This obstacle can be removed if one takes into account one of the most fundamental and general theorems valid in thermodynamics of irreversible processes: the General Evolution Criterion. Glansdorff and Prigogine have shown that: \textit{For time‐independent boundary conditions, a thermodynamic system, even in strong non‐equilibrium conditions, relaxes to a stable stationary state in such a way that the following General Evolution Criterion is satisfied} 
\begin{equation}\label{RDDP1}
P=\int_\Omega J_\mu\frac{\partial X^\mu}{\partial t} d{\rm v}\leq 0
\end{equation}
\noindent In addition 
\begin{equation}\label{RDDP2}
P=\int_\Omega J_\mu\frac{\partial X^\mu}{\partial t} d{\rm v} = 0 \quad{\rm at\ the\ steady\ state}
\end{equation}
\noindent Quantity $P$ may be referred to as the \textit{Glansdorff‐Prigogine dissipative quantity}. Let us check the validity of this theorem by considering two, very simple, examples. Let us consider, for instance, a closed system containing $m$ components ($i=1\cdots m$) among which chemical reactions are possible. The temperature, $T$, and the pressure, $p$, are supposed to be constant in time. The chance in the number of moles $n_i$, of component $i$, is 
\begin{equation}\label{RDDP3}
\frac{dn_i}{dt} = \nu^j_i v_j
\end{equation}
\noindent with $\nu_i^j$ denoting the \textit{stoichiometric coefficients}. By multiplying both members of Eq.~(\ref{RDDP3}) by the time derivative of the chemical potential of component $i$, we get
\begin{equation}\label{RDDP4}
\frac{d\mu^i}{dt}\frac{dn_i}{dt} = \Bigl(\frac{\partial \mu^i}{\partial n_\iota}\Bigr)_{(pT)}\frac{dn_i}{dt} \frac{dn_\iota}{dt} =\Bigl(\frac{\partial \mu^i}{\partial n_\iota}\Bigr)_{(pT)}\nu^j_i\nu^\kappa_\iota v_j v_\kappa\geq 0
\end{equation}
\noindent where the positive sign of the term on the right-hand side is due to the second law of thermodynamics. By taking into account the De Donder law between the affinities $A^j$ and the chemical potentials i.e., $A^j=-\nu_i^j \mu^i$, and that the chemical thermodynamic force ($X^j$) and its chemical conjugate flux ($J_j$) read $X^i=A^j/T$ and $J_i=v_j$ respectively, we finally get
\begin{equation}\label{RDDP4}
P\!=\!\!\int_\Omega \! J_\mu\frac{\partial X^\mu}{\partial t} d{\rm v}\! =\Omega J_\mu\frac{dX^\mu}{dt}\!=\Omega v_j\frac{d}{dt} \Bigl(\frac{A^j}{T}\Bigr)=-\frac{\Omega}{T}\Bigl(\frac{\partial\mu^i}{\partial n_\iota}\Bigr)_{(pT)}\!\nu^j_i\nu^\kappa_\iota v_j v_\kappa\leq 0
\end{equation}
\noindent Hence, the Glansdorff-Prigogine dissipative quantity $P$ is always negative and it vanishes at the stationary state. As a second example, we analyze the case of heat conduction in non‐expanding solid. In this case the thermodynamic forces and the conjugate fluxes are the (three) components of the gradient of the inverse of the temperature, $X^\mu= {\boldsymbol \nabla}(1/T)$, and the (three) components of the heat flux, $J_\mu={\boldsymbol J}_{(q)}$ (with $\mu=1,2,3$), respectively. Hence,
\begin{equation}\label{RDDP5}
P=\int_\Omega \! J_\mu\frac{\partial X^\mu}{\partial t} d{\rm v}=\int_\Omega {\boldsymbol  J}_{(q)}\cdot {\boldsymbol \nabla} (1/T) d{\rm v}
\end{equation}
\noindent The heat flux, ${\boldsymbol J}_{(q)}$, is linked to the (partial) time derivative of temperature by the Fourier law (expressing the energy balance equation)
\begin{equation}\label{RDDP6}
\rho c_v\frac{\partial T}{\partial t}= -{\boldsymbol \nabla} \cdot {\boldsymbol J}_q
\end{equation}
\noindent with $\rho$ and $c_v$ denoting the mass density and the specific heat at volume constant of the fluid, respectively. By performing the integration by parts, and by assuming that the heat flux vanishes at the boundary, we easily get
\begin{equation}\label{RDDP7}
P=-\int_\Omega\frac{\rho c_v}{T^2}\Bigl(\frac{\partial T}{\partial t}\Bigr)^2 d{\rm v}\leq 0
\end{equation}
\noindent with $P=0$ at the steady state. By summarising, \textit {for all thermodynamic systems, without using the Onsager reciprocal relations, and even if the transport coefficients dependent on the thermodynamic forces, the dissipative quantity} $P$ {\textit{is always a negative quantity}. This quantity vanishes at the steady-state. In the two above‐mentioned examples, the thermodynamic forces are the chemical affinities (over temperature) and the gradient of the inverse of temperature, respectively. However, we could have adopted a different choice of the thermodynamic forces. If we analyze, for instance, the case of heat conduction in non‐expanding solid, where chemical reactions take place simultaneously, we can choose as thermodynamic forces a combination of the (dimensionless) chemical affinities (over temperature) and the (dimensionless) gradient of the inverse of temperature. Clearly, this representation is thermodynamically equivalent to the previous one (where the thermodynamic forces are simply the chemical affinities over temperature and the gradient of the inverse of temperature) only if the negative sign of the dissipative quantity $P$ is preserved. In other words, the equations providing the stationary states (i.e., Eq.~(\ref{RDDP2})) must admit exactly the same solutions.

\subsection{The Thermodynamic Covariant Transformations (TCT) and the Thermodynamic Covariance Principle (TCP)}\label{TCP}
One of the central aspects of the TFT is the concept of invariance of physics' laws. This invariance can be described in many ways, for example, in terms of local covariance or covariance of diffeomorphism. A more explicit description can be given through the use of tensors. The characteristic of the tensors that proves to be crucial is the fact that, once given the metric, the operation of contracting a tensor of rank $r$ on all indices $r$ provides a number - an {\it invariant} - which is independent of the coordinates used to perform the contraction. Physically, this means that the invariant calculated by choosing a specific coordinate system (i.e., in a specific set  of the thermodynamic forces) will have the same value if calculated in another - thermodynamically equivalent - coordinate system (i.e. in another equivalent set of thermodynamic forces). \textit{According to the Thermodynamical Field Theory} (TFT), \textit{two set of thermodynamic forces are equivalent if the following two conditions are satisfied} \footnote{According to the TFT conditions {\bf i)} and {\bf ii)} establish the {\it equivalent character} between two different representations (i.e., between two different set of thermodynamic forces.)} \cite{sonnino1}:

\noindent (i) The entropy production $\sigma$ must be invariant under transformation of the thermodynamic forces $\{X^\mu\}\rightarrow\{X^{'\mu}\}$.

\noindent (ii) The Glansdorff-Prigogine dissipative quantity $P$ must also be invariant under the force transformations $\{X^\mu\}\rightarrow \{X^{'\mu}\}$.

\noindent Condition ii) stems from the fact that a stable steady-state must be transformed into the same stable-state state, with the \textit{same degree of stability}. In mathematical terms, this implies
\begin{equation}\label{TCP1}
\sigma=J_\mu X^\mu = J'_\mu X^{'\mu}=\sigma'\ ;\ P=P'\rightarrow J_\mu\delta X^\mu=J'_\mu\delta X^{'\mu}\qquad{\rm and}\quad t=t'
\end{equation}
\noindent Eqs.~(\ref{TCP1}) are satisfied iff the transformed thermodynamic forces and conjugate fluxes read as \cite{sonnino1}, \cite{sonnino4}, \cite{sonnino5}
\begin{equation}\label{TCP2}
X^{'\mu}=\frac{\partial X^{'\mu}}{\partial X^\nu}X^\nu\quad ,\quad J'_\mu=\frac{\partial X^\nu}{\partial X^{'\mu}}J_\nu
\end{equation}
\noindent Transformations~(\ref{TCP2}) are referred to as the {\it Thermodynamic Covariant Transformations} (TCT) \cite{sonnino1}. The thermodynamic equivalence principle leads, naturally, to the following {\it Thermodynamic Covariance Principle} (TCP) \cite{sonnino4}, \cite{sonnino5}: 

\textit{The nonlinear closure equations, i.e. the flux-force relations, must be covariant under} TCT. 

\noindent The essence of the TCP is the following. The equivalent character between two representations is warranted iff the fundamental thermodynamic equations (e.g., the transport equations) are covariant under the Thermodynamic Covariant Transformations (TCT).

\subsection{The TCT-Symmetry Group}\label{TCTgroup}
\subsubsection{Topological Structure the TCT-Group}
The invariance of a system under TCP is intimately related to the existence of a group, which we refer to as the {\it TCT-group} \cite{sonnino6}, \cite{sonnino7}. The TCT-group with its properties can be identified by analysing the solution of Eq.~(\ref {TCP2}). The solution of Eq.~(\ref{TCP2}) reads 
\begin{equation}\label{TCTgroup1}
X^{'\mu}=X^1F^\mu\left(\frac{X^2}{X^1}, \frac{X^3}{X^2},\cdots \frac{X^n}{X^{n-1}}\right)
\end{equation}
\noindent with $F^{\mu}$ denoting arbitrary functions. Hence, the ratio $\{X^\mu/X^{\mu-1}\}$ are the coordinates for a different space: the {\it Real Projective Space} ${R}{P}^{n-1}$, which is defined to be the quotient of ${R}^n$ minus the origin by the scaling map $X^\mu\rightarrow \alpha X^\mu$ with $\alpha$ denoting any nonzero real number - see Fig.~(\ref{TFT_Projective_plane}). The TCT-group is then {\it the product of} ${\rm diff} ({\mathbb R}{\mathbb P}^{n-1})$ {\it with the multiplicative group of the map from} ${\mathbb R}{\mathbb P}^{n-1}\rightarrow {\mathbb R}^\times$ - see Fig.~(\ref{TCP_Group2}) \cite{sonnino6}, \cite{sonnino7}.

\begin{figure}[b]
\sidecaption
\includegraphics[scale=.65]{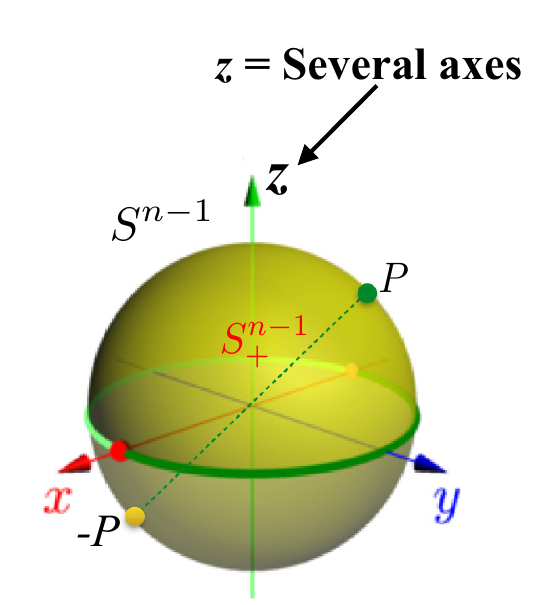}
\caption{{\bf The projective space}. The Projective space ${\mathbb R}{\mathbb P}^{n-1}$ is diffeomophic to $S_+^{n-1}$ made by the upper hemisphere + half equator (without the red and yellow points) + the red point.}\label{TFT_Projective_plane}
\end{figure}

\begin{figure}[b]
\sidecaption
\includegraphics[width=6cm]{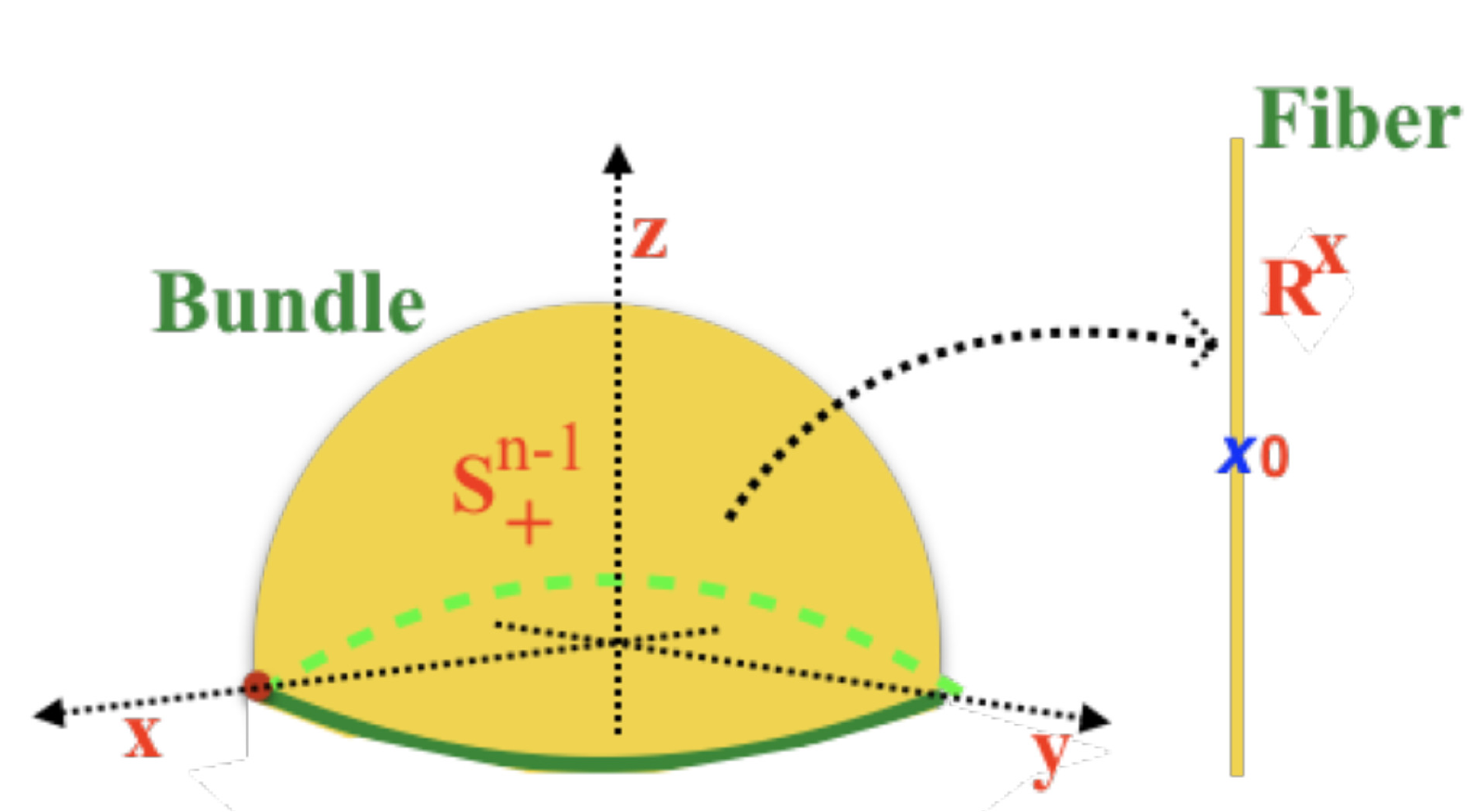}
\caption{{\bf The TCT-group}. The TCT-group $G^n$ is the application $F^\mu$, from the {\it bundle} (the projective space ${\mathbb R}{\mathbb P}^{n-1}$) to the {\it fiber} (${\mathbb R}^x$).}
\label{TCP_Group2}
\end{figure}

\subsubsection{Algebraic Structure of the TCT-Group}\label{ATCTgroup}
In the previous Subsection, we have seen the the TCT group, denoted by $G^n$, is a specific subgroup of the homogeneous diffeomorphisms from ${\rm diff}({\mathbb R}\!\!\setminus\!\! (\{0\})$. In algebraic terms, the result of the previous Subsection may be expressed as follows: {\it The $G^n$ results from the application}: 

\noindent {\it ${\rm diff}({\mathbb R}\!\!\setminus\!\! (\{0\})\ni X\vdash Y_g(X)\in{\rm diff}({\mathbb R}\!\!\setminus\!\! (\{0\})$}, {\it with} $Y_g\in G^n$ iff $Y_g(\lambda X)=\lambda Y_g(X)$ with $\lambda\in{\mathbb R}$.

\noindent It is possible to demonstrate that the TCT-group, $G^n$, may be split in a semidirect product of two subgroups where the first one is a {\it normal}, Abelian, subgroup, denoted by $N^n$, and the second one is the {\it reflection} subgroup. The demonstration of this theorem can be found in ref.~\cite{sonnino6}. More specifically, let us introduce two subgroups $N^n$ and $H^n$ defined as follows. Let $N^n$ denote the {\it normal subgroup} of $G^n$ defined as
\begin{equation}\label{ATCTgroup1}
N^n: \ Y_g(X)=X r_g(X)\ {\rm with}\ r_g(\lambda X)=r_g(X) >0
\end{equation} 
\noindent with $\lambda >0$. Here, $r_g(X)$ is a positive ${\mathbb C}^\infty({\mathbb R}\!\!\setminus\!\! (\{0\}))$ homogeneous function.

\noindent Let $H^n$ denote the {\it reflection} subgroup of $G^n$ defined as 
\begin{align}\label{ATCTgroup2}
&H^n:\ \parallel Y_h({\bf X})\parallel=\parallel X\parallel\  ;\  Y_h(-X)=-Y_h(X)\\
&{\rm with}\ h\in H^n\nonumber
\end{align} 
\noindent In ref.~\cite{sonnino6} it is proved that
\begin{equation}\label{ATCTgroup3}
G^n=N^n\rtimes H^n
\end{equation} 
\noindent The irreducible representations of the group G are then related to the irreducible representations of the subgroups $N^n$ and $H^n$.

\subsection{The Thermodynamic Action Principle}\label{TAP}
Constraint 2. and assumption 3., reported in the Introduction, lead to the following \textit{thermodynamic action principle} \cite{sonnino1}: 
\begin{itemize}
\item{\textit{There exists a thermodynamic action $I$, scalar under Thermodynamic Covariant Transformations (TCT), which is stationary with respect to arbitrary variations in the transport coefficients and the affine connection}}. 
\end{itemize}
\noindent This action, scalar under TCT, must be constructed only by the transport coefficients, the affine connection, and their first derivatives. In addition, it must be linear in the second derivatives of the transport coefficients and it cannot contain second or higher derivatives of the affine connection. We also require that the action is stationary when the affine connection takes the following expression \cite{sonnino1}:
\begin{equation}\label{TAP1}
\Gamma^\lambda_{\mu\nu}={\widetilde\Gamma}^\lambda_{\mu\nu}
\end{equation}
\noindent Hence, our Lagrangian density ${\mathcal L}$ depends on three sets of dynamical variables: ${\mathcal L}={\mathcal L}(g_{\mu\nu},\ f_{\mu\nu},\ \Gamma^\lambda_{\mu\nu})$. The simplest action satisfying these requirements is
\begin{equation}\label{TAP2}
I=\int{\!\!\mathcal L}\ \sqrt{g}d^nX=\int\left[B-(\Gamma^\lambda_{\mu\nu}-{\widetilde\Gamma}^\lambda_{\alpha\beta})S^{\mu\nu}_\lambda + {\widetilde{\mathcal L}}(g_{\mu\nu},f_{\mu\nu})\right]\sqrt{g}d^nX
\end{equation}
\noindent with $B$ denoting the \textit{scalar curvature of the thermodynamic space}\footnote{To avoid misunderstanding with the Riemannian (or Pseudo-Riemannian) geometry, we adopt the Eisenhart notations \cite{eisenhart}.}:
\begin{equation}\label{TAP3}
B=B_{\mu\nu}g^{\mu\nu}\quad ;\quad B_{\mu\nu}=\frac{\partial\Gamma^\lambda_{\mu\lambda}}{\partial X^\nu}-\frac{\partial\Gamma^\lambda_{\mu\nu}}{\partial X^\lambda}+\Gamma^\eta_{\mu\lambda}\Gamma^\lambda_{\eta\nu}-\Gamma^\eta_{\mu\nu}\Gamma^\lambda_{\eta\lambda}
\end{equation}
\noindent and the expressions of $S^{\alpha\beta}_\lambda$ is \cite{sonnino1}:
\begin{equation}\label{TAP4}
S^{\mu\nu}_\lambda=\Psi^\mu_{\lambda\alpha}g^{\nu\alpha}+\Psi^\nu_{\lambda\alpha}g^{\mu\alpha}-\frac{1}{2}\Psi^\mu_{\alpha\beta}g^{\alpha\beta}\delta_\lambda^\nu-\frac{1}{2}\Psi^\nu_{\alpha\beta}g^{\alpha\beta}\delta^\mu_\lambda
\end{equation}
\noindent with
\begin{eqnarray}\label{TAP5}
&&\Psi^\mu_{\lambda\kappa}=\frac{1}{2}\tau^{\mu\eta}g_{\lambda\kappa,\eta}-\frac{1}{2(n+1)}\tau^{\eta\alpha}g_{\lambda\eta,\alpha}\delta^\mu_\kappa-\frac{1}{2(n+1)}\tau^{\eta\alpha}g_{\kappa\eta,\alpha}\delta^\mu_\lambda\\
&&\tau^{\mu\nu}=\frac{X^\mu X^\nu}{\sigma}\quad ;\quad \sigma=g_{\mu\nu}X^\mu \!X^\nu\nonumber
\end{eqnarray}
\noindent ${\widetilde{\mathcal L}}(g_{\mu\nu},f_{\mu\nu})$ is a Lagrangian density that may depend on the transport coefficients but not on the affine connection. Note that $\tau^{\mu\nu}$ is a second-rank thermodynamic tensor. The physical meaning of the Lagrangian density stems from its (strict) connection with the curvature of the thermodynamic space. So, we require that the Lagrangian density must coincide with the scalar curvature $B$ when the affine connection takes the expression ${\widetilde\Gamma}^\lambda_{\mu\nu}$. This is because the scalar $B$ is the simplest curvature scalar, and the only one that is linear in the curvature of the space. This implies that $ {\widetilde{\mathcal L}}\equiv 0$ and the final expression of the thermodynamic action reads
\begin{equation}\label{TAP6}
I=\int\left[B-(\Gamma^\lambda_{\mu\nu}-{\widetilde\Gamma}^\lambda_{\mu\nu})S^{\mu\nu}_\lambda\right]\sqrt{g}d^n X
\end{equation}

\subsection{The Privileged Thermodynamic Coordinate System}\label{PTCS}
By definition, a \textit{thermodynamic coordinate system is a complete set of independent thermodynamic forces}. Once a particular set of thermodynamic coordinates is choosen, the other sets of coordinates are linked to the first one through a {\it Thermodynamic Coordinates Transformation} (TCT). The simplest way to determine a particular set of coordinates is to quote the entropy balance equation
\begin{equation}\label{P1}
\frac{\partial \rho s}{\partial t}+{\bf\nabla}\cdot{\bf J}_s=\sigma
\end{equation}
 \noindent Here, $\rho s$ is the local total entropy per unit volume ($\rho$ is the mass density) and ${\bf J}_s$ is the entropy flux, respectively. Let us consider, as an example, a thermodynamic system confined in a rectangular box where chemical reactions, diffusion of matter, macroscopic motion of the volume element (convection) and heat current take place simultaneously. The entropy flux and the entropy production read \cite{degroot}, \cite{fitts}, \cite{vidal}
 \begin{eqnarray}\label{P2}
&&\!\!\!\!\!\!\!\!\!\!\!\!\!\!\!\!\!\!\!\!
 {\bf J}_s=\frac{1}{T}({\bf J}_q-\sum_i {\bf J}_i\mu_i)+\sum_i\rho_i v_is_i\\
&&\!\!\!\!\!\!\!\!\!\!\!\!\!\!\!\!\!\!\!\!
\sigma={\bf J}_q\!\cdot\!{\bf\nabla}\frac{1}{T}\!-\!\frac{1}{T}\!\sum_i{\bf J}_i\!\cdot\!\Bigl[T{\bf\nabla}\Bigl(\frac{\mu_i}{T}\Bigl)\!-\!{\bf F}_i\Bigr]\!+\!\sum_i\frac{w_iA_i}{T}\!-\!\frac{1}{T}\!\sum_{ij}\Pi_{ij}\partial_{{\bf r}_i}v_j\geq0\nonumber
\end{eqnarray}
 \noindent where $\mu_i$, $\rho_i s_i$ and $A_i$ are the chemical potential, the local entropy and the affinity of species $"i"$, respectively. ${\bf J}_q$ is the heat flux; ${\bf J}_i$ and $w_i$ are the diffusion flux and the chemical reaction rate of species $i$, respectively. Moreover, $\Pi_{ij}$ are the components of the dissipative part of the pressure tensor ${\mathcal M}_{ij}$ (${\mathcal M}_{ij}=p\delta_{ij}+\Pi_{ij}$; $p$ is the hydrostatic pressure), ${\bf F}_i$ the external force per unit mass acting on $i$, and $v_j$ denotes the component of the hydrodynamic velocity \cite{vidal}. The set of the thermodynamic coordinates reads
 \begin{equation}\label{P3}
X^\mu=\left\{{\bf\nabla}\left(\frac{1}{T}\right);\ -\frac{1}{T}\left(T{\bf\nabla}\left(\frac{\mu_i}{T}\right)-{\bf F}_i\right);\ \frac{A_i}{T};\ -\frac{1}{T}\partial_{{\bf r}_i}v_j\right\}
\end{equation}
 \noindent For this particular example, this set may be referred to as the \textit{privileged thermodynamic coordinates system}. Other examples of privileged thermodynamic coordinates system, related to magnetically confined plasmas, can be found in refs. \cite{balescu2}, \cite{balescu1}, and \cite{hinton}.

\section{Transport Equations}\label{nte}
Action~(\ref{TAP6}) is stationary with respect to small, and arbitrary, variations of the {\it dynamical variable} $g_{\mu\nu}$ and $\Gamma^\lambda_{\mu\nu}$ (we set $f_{\mu\nu}=0$). We recall that action ~(\ref{TAP6}) has been constructed in such a way that it is stationary for $\Gamma^\lambda_{\mu\nu}={\widetilde\Gamma}^\lambda_{\mu\nu}$. Indeed, by variational calculations, we get that the action is stationary with respect to small, independent, variations of $g_{\mu\nu}$ and $\Gamma^\lambda_{\mu\nu}$ if \cite{sonnino1}
\begin{eqnarray}\label{nte1}
&&B_{\mu\nu}-\frac{1}{n-2}\ g_{\mu\nu}B=-S^{\lambda\kappa}_\eta\frac{\delta{\widetilde\Gamma}^\eta_{\lambda\kappa}}{\delta g^{\mu\nu}}\equiv T_{\mu\nu}\\
&&\Gamma^\lambda_{\mu\nu}={\widetilde\Gamma}^\lambda_{\mu\nu}\nonumber
\end{eqnarray}
\noindent  Eq.~(\ref{nte1}) is valid for $n\neq 2$. Much less easy is to compute the explicit expression of $T_{\mu\nu}$. After (quite long) calculations, we get
\begin{eqnarray}\label{nte2}
{T}_{\mu\nu}&\equiv& -S^{\lambda\kappa}_\eta\frac{\delta{\widetilde\Gamma}^\eta_{\lambda\kappa}}{\delta g^{\mu\nu}}=\\
&&\frac{1}{2}g^{-1/2}(S_\alpha^{\lambda\kappa}g^{\alpha\beta}g^{1/2})_{,\beta}g_{\lambda\mu}g_{\kappa\nu}-\frac{1}{2}g^{-1/2}\Bigl(g^{1/2}(S_\mu^{\kappa\beta}g_{\kappa\nu}+S_\nu^{\kappa\beta}g_{\kappa\mu})\Bigr)_{,\beta}\nonumber\\
& -&\frac{1}{4}S_\mu^{\eta h}(g_{\nu\eta,h}+g_{\nu h,\eta}-g_{\eta h,\nu})-\frac{1}{4}S_\nu^{\eta h}(g_{\mu\eta,h}+g_{\mu h,\eta}-g_{\eta h,\mu})\nonumber\\
&-&\frac{1}{2}g^{-1/2}\left(g^{1/2}\tau^{\eta\beta}S_\eta^{\lambda\kappa}-\frac{g^{1/2}}{n+1}\left(\tau^{\lambda\beta}S^{\eta\kappa}_\eta+\tau^{\kappa\beta}S^{\eta\lambda}_\eta\right)\right)_{,\beta}g_{\lambda\mu}g_{\kappa\nu}\nonumber\\
&-&\frac{1}{2}\tau^{\alpha\beta}\tau^{\lambda\kappa}S^{\eta\iota}_\alpha g_{\eta\iota ,\beta}g_{\lambda\mu}g_{\kappa\nu}+\frac{1}{n+1}\tau^{\alpha\beta}\tau^{\lambda\kappa}S^{\eta\iota}_\eta g_{\alpha\iota ,\beta}g_{\lambda\mu}g_{\kappa\nu}
\end{eqnarray}
\noindent After a little algebra, we find
\begin{eqnarray}\label{nte3}
&-&\frac{1}{2}g^{-1/2}\left(g^{1/2}\tau^{\eta\beta}S_\eta^{\lambda\kappa}-\frac{g^{1/2}}{n+1}\left(\tau^{\lambda\beta}S^{\eta\kappa}_\eta+\tau^{\kappa\beta}S^{\eta\lambda}_\eta\right)\right)_{,\beta}g_{\lambda\mu}g_{\kappa\nu} =\\
&-&\frac{1}{2}\left(\Psi^\lambda_{\eta\alpha}g^{\kappa\alpha}_{,\beta}+\Psi^\kappa_{\eta\alpha}g^{\lambda\alpha}_{, \beta}\right)\tau^{\eta\beta}g_{\lambda\mu}g_{\kappa\nu}\nonumber\\
&+&\frac{1}{2(n+1)}g^{-1/2}\left(g^{1/2}\left(\Psi^\lambda_{\eta\alpha}g^{\eta\alpha}\tau^{\kappa\beta}+\Psi^\kappa_{\eta\alpha}g^{\eta\alpha}\tau^{\lambda\beta}\right)\right)_{, \beta}g_{\lambda\mu}g_{\kappa\nu}\nonumber\\
&-&\frac{1}{2}g^{-1/2}g_{\kappa\mu}\left(g^{1/2}\Psi^\kappa_{\eta\nu}\tau^{\eta\beta}\right)_{, \beta}-\frac{1}{2}g^{-1/2}g_{\kappa\nu}\left(g^{1/2}\Psi^\kappa_{\eta\mu}\tau^{\eta\beta}\right)_{, \beta}\nonumber
\end{eqnarray}
\noindent where we note that the trace of the last expression of Eq.~(\ref{nte3}) vanishes. We also have
\begin{eqnarray}\label{nte4}
&-&\frac{1}{2}\tau^{\alpha\beta}\tau^{\lambda\kappa}S^{\eta\iota}_\alpha g_{\eta\iota ,\beta}g_{\lambda\mu}g_{\kappa\nu}+\frac{1}{n+1}\tau^{\alpha\beta}\tau^{\lambda\kappa}S^{\eta\iota}_\eta g_{\alpha\iota ,\beta}g_{\lambda\mu}g_{\kappa\nu} =\\
&-&\Psi^\eta_{\alpha\gamma}g^{\iota\gamma}g_{\eta\iota ,\beta}\tau^{\alpha\beta}\tau^{\lambda\kappa}g_{\lambda\mu}g_{\kappa\nu}+\frac{1}{n+1}\Psi^\iota_{\eta\gamma}g^{\eta\gamma}g_{\alpha\iota ,\beta}\tau^{\alpha\beta}\tau^{\lambda\kappa}g_{\lambda\mu}g_{\kappa\nu}\nonumber
\end{eqnarray}
\noindent Hence, tensor $T_{\mu\nu}$ can be brought into the form
\begin{eqnarray}\label{nte5}
{T}_{\mu\nu}&=&\frac{1}{2}g^{-1/2}(S_\alpha^{\lambda\kappa}g^{\alpha\beta}g^{1/2})_{,\beta}g_{\lambda\mu}g_{\kappa\nu}-\frac{1}{2}g^{-1/2}\Bigl(g^{1/2}(S_\mu^{\kappa\beta}g_{\kappa\nu}+S_\nu^{\kappa\beta}g_{\kappa\mu})\Bigr)_{,\beta}\nonumber\\
& -&\frac{1}{4}S_\mu^{\eta h}(g_{\nu\eta,h}+g_{\nu h,\eta}-g_{\eta h,\nu})-\frac{1}{4}S_\nu^{\eta h}(g_{\mu\eta,h}+g_{\mu h,\eta}-g_{\eta h,\mu})\nonumber\\
&-&\frac{1}{2}g^{-1/2}\left(g^{1/2}\tau^{\eta\beta}S_\eta^{\lambda\kappa}-\frac{g^{1/2}}{n+1}\left(\tau^{\lambda\beta}S^{\eta\kappa}_\eta+\tau^{\kappa\beta}S^{\eta\lambda}_\eta\right)\right)_{,\beta}g_{\lambda\mu}g_{\kappa\nu}\nonumber\\
&-&\Psi^\eta_{\alpha\gamma}g^{\iota\gamma}g_{\eta\iota ,\beta}\tau^{\alpha\beta}\tau^{\lambda\kappa}g_{\lambda\mu}g_{\kappa\nu}+\frac{1}{n+1}\Psi^\iota_{\eta\gamma}g^{\eta\gamma}g_{\alpha\iota ,\beta}\tau^{\alpha\beta}\tau^{\lambda\kappa}g_{\lambda\mu}g_{\kappa\nu}
\end{eqnarray}
\noindent and the trace of tensor $T_{\mu\nu}$ reads
\begin{eqnarray}\label{nte6}
{T}\equiv {T}_{\mu\nu}g^{\mu\nu}&=&\frac{n-2}{2}g^{-1/2}(g^{1/2}\Psi_{\lambda\kappa}^\beta g^{\lambda\kappa})_{,\beta}+\frac{1}{n+1}g_{\eta\alpha}g^{-1/2}(g^{1/2}\Psi_{\lambda\kappa}^\alpha g^{\lambda\kappa}\tau^{\eta\beta})_{,\beta}\nonumber\\
&-&\Psi_{\alpha\gamma}^\eta\tau^{\alpha\beta}\tau^{\lambda\kappa}g^{\iota\gamma}g_{\lambda\kappa}g_{\eta\iota ,\beta}+\frac{1}{n+1}\Psi_{\eta\gamma}^\iota g^{\eta\gamma}\tau^{\alpha\beta}\tau^{\lambda\kappa}g_{\lambda\kappa} g_{\alpha\iota ,\beta}\nonumber\\
&+&2S_\eta^{\kappa\beta}g_{,\beta}^{\eta\lambda}g_{\lambda\kappa}-\Psi^\kappa_{\alpha\eta}\tau^{\alpha\beta}g_{\lambda\kappa}g^{\lambda\eta}_{,\beta}
\end{eqnarray}
\noindent Finally, \emph{in absence of the skew-symmetric part, we get the differential equations for the transport coefficients valid for} $n >2$:
\begin{equation}\label{nte7}
B_{\mu\nu}=T_{\mu\nu}-\frac{1}{n-2}\ g_{\mu\nu}T=W^{(S)}_{\mu\nu}
\end{equation}
\noindent with $T_{\mu\nu}$ and $T$ given by Eq.~(\ref{nte5}) and Eq.~(\ref{nte6}), respectively.
\vskip0.2truecm
\noindent {\bf $\bullet$ Property of Tensor $W^{(S)}_{\mu\nu}$}

\noindent Tensor $B_{\mu\nu}$ satisfies the Bianchi identity for symmetric connection which, written in the linearised form, reads
\begin{equation}\label{nte8}
L^{\mu\lambda}\frac{\partial B_{\lambda\nu}}{\partial X^\mu}-\frac{1}{2}L^{\mu\lambda}\frac{\partial B_{\mu\lambda}}{\partial X^\nu}\equiv 0
\end{equation}
\noindent The validity of identity~(\ref{nte8}) may also be checked by direct inspection. Hence, also tensor $W^{(S)}_{\mu\nu}(h)$ satisfies the same identity
\begin{equation}\label{nte9}
L^{\mu\lambda}\frac{\partial W^{(S)}_{\lambda\nu}}{\partial X^\mu}-\frac{1}{2}L^{\mu\lambda}\frac{\partial W^{(S)}_{\mu\lambda}}{\partial X^\nu}\equiv 0
\end{equation}
\vskip0.2truecm
\noindent {\bf $\bullet$ Observations}

\noindent By direct inspection, we may check the validity of the following important identities

\begin{equation}\label{nte10}
\Psi^\lambda_{\lambda\kappa}=\Psi^\lambda_{\kappa\lambda}=0\quad ;\quad S_\lambda^{\lambda\mu}=-\frac{(n-1)}{2}\Psi_{\lambda\kappa}^\mu g^{\lambda\kappa}
\end{equation}

\subsection {Onsager's Region}\label{OR}

\noindent The transport coefficients tend to Onsager's matrix as the thermodynamic system approaches thermodynamic equilibrium. The thermodynamic region where the thermodynamic forces are linearly connected to the conjugate thermodynamic fluxes is referred to as the {\it linear region of thermodynamics} or {\it Onsager's region} \cite{prigogine1}, \cite{prigogine2}. Hence, as the thermodynamic forces go to zero, the metric $g_{\mu\nu}$ tends to Onsager matrix $L_{\mu\nu}$ (or, equivalently, the perturbation $h_{\mu\nu}$ of the metric tensor tends to zero):
\begin{equation}\label{OR1}
\lim_{X^\lambda\rightarrow 0}g_{\mu\nu}=L_{\mu\nu}
\end{equation}
\noindent Condition~(\ref{OR1}) is referred to as {\it Onsager's condition}.

\subsection{Near the Onsager Region}\label{NLR}
Let us compute the first nonlinear contributions of Eq.~(\ref{nte7}). In this case, since $\Psi^\lambda_{\mu\nu}$ is already of the first order in $h_{\mu\nu}$, to obtain equations valid up to the third order, we should develop quantities $g^{\mu\nu}$, $g$, $\sigma^2$, $\tau^{\lambda\kappa}$ etc. up to the second order. Hence, by setting
\begin{equation}\label{nl1}
g_{\mu\nu}\simeq L_{\mu\nu}+h_{\mu\nu}\quad ;\quad \sigma\simeq \sigma_{(L)}(1+h_{\mu\nu}\tau_{(L)}^{\mu\nu})
\end{equation}
\noindent with $\sigma_{(L)}\equiv L_{\mu\nu}X^\mu X^\nu$ and $\tau_{(L)}^{\mu\nu}\equiv X^\mu X^\nu/\sigma_{(L)}$, we get
\begin{eqnarray}\label{nl2}
g^{\mu\nu}&\simeq& L^{\mu\nu}-h^{\mu\nu}+h^{\mu\lambda}h^{\nu\kappa}L_{\lambda\kappa}+O(h^3)\\
\tau^{\mu\nu}&\simeq&\tau_{(L)}^{\mu\nu}\left(1-h_{\lambda\kappa}\tau_{(L)}^{\lambda\kappa}+(h_{\lambda\kappa}\tau_{(L)}^{\lambda\kappa})^2\right)+O(h^3)
\nonumber\\
\sigma^2&\simeq&\sigma_{(L})\left(1+2h_{\mu\nu}\tau_{(L)}^{\mu\nu}+(h_{\mu\nu}\tau_{(L)}^{\mu\nu})^2\right)+O(h^3)
\nonumber\\
\frac{1}{\sigma}&\simeq& \frac{1}{\sigma_{(L)}}\left(1-h_{\lambda\kappa}\tau^{\lambda\kappa}_{(L)}+(h_{\lambda\kappa}\tau^{\lambda\kappa}_{(L)})^2\right)+O(h^3)\nonumber\\
\frac{1}{\sigma^2}&\simeq& \frac{1}{\sigma^2_{(L)}}\left(1-2h_{\lambda\kappa}\tau^{\lambda\kappa}_{(L)}+3(h_{\lambda\kappa}\tau^{\lambda\kappa}_{(L)})^2\right)+O(h^3)
\nonumber\\
g&\simeq& L\Bigl(1+L^{\lambda\kappa} h_{\lambda\kappa}+\frac{1}{2}\bigl((L^{\lambda\kappa}h_{\lambda\kappa})^2-L^{\lambda\alpha}L^{\kappa\beta}h_{\alpha\kappa}h_{\beta\lambda}\bigr)\Bigr)+O(h^3)\nonumber\\
g^{1/2}&\simeq& L^{1/2}\Bigl(1+\frac{1}{2}L^{\lambda\kappa} h_{\lambda\kappa}+\frac{1}{8}(L^{\lambda\kappa}h_{\lambda\kappa})^2-\frac{1}{4}L^{\lambda\alpha}L^{\kappa\beta}h_{\alpha\kappa}h_{\beta\lambda}\Bigr)+O(h^3)\nonumber\\
g^{-1/2}&\simeq& L^{-1/2}\Bigl(1-\frac{1}{2}L^{\lambda\kappa} h_{\lambda\kappa}+\frac{1}{8}(L^{\lambda\kappa}h_{\lambda\kappa})^2+\frac{1}{4}L^{\lambda\alpha}L^{\kappa\beta}h_{\alpha\kappa}h_{\beta\lambda}\Bigr)+O(h^3)\nonumber
\end{eqnarray}
\noindent with $L$ denoting the determinant of Onsager's matrix.

\section{Two-Dimensional Transport Equations}\label{2dim}
In two dimensions, the curvature tensor $B_{\mu\nu\lambda\kappa}$ has only one component, since all nonzero components may be obtained from $B_{0101}$. Equivalently, the curvature tensor may be written in terms of the scalar $B$
\begin{equation}\label{2dim1}
B_{\lambda\mu\kappa\nu}=\frac{1}{2}B\left(g_{\lambda\kappa}g_{\mu\nu}-g_{\lambda\nu}g_{\mu\kappa}\right)
\end{equation}
\noindent So, $B$ alone completely characterises the local geometry. From Eq.~(\ref{2dim1}) we find the expressions for $B_{\mu\nu}\equiv B_{\lambda\mu\kappa\nu}g^{\lambda\kappa}$ and $B\equiv B_{\lambda\mu\kappa\nu}g^{\lambda\kappa}g^{\mu\nu}$. We get
\begin{equation}\label{2dim2}
B_{\mu\nu}-\frac{1}{2}Bg_{\mu\nu}\equiv 0
\end{equation}
\noindent Hence, Eq.~(\ref{nte7}) is meaningless in two dimensions (see also refs.~\cite{brown}, \cite{collas}). It is easy to convince ourselves that, in analogy with the works for $1+1$ gravity \cite{jackiw}, \cite{teitelboim}, also in our case the only non-trivial version of the Eq.~(\ref{nte7}) for $n=2$ has to read
\begin{equation}\label{2dim3}
B=-T
\end{equation}
\noindent with
\begin{eqnarray}\label{2dim4}
T=&&2S_\eta^{\kappa\beta}g_{,\beta}^{\eta\lambda}g_{\lambda\kappa}\\
&-&\Psi^\kappa_{\alpha\eta}\tau^{\alpha\beta}g_{\lambda\kappa}g^{\lambda\eta}_{,\beta}-\Psi_{\alpha\gamma}^\eta\tau^{\alpha\beta}\tau^{\lambda\kappa}g^{\iota\gamma}g_{\lambda\kappa}g_{\eta\iota ,\beta}\nonumber\\
&+&\frac{1}{3}g_{\eta\alpha}g^{-1/2}(g^{1/2}\Psi_{\lambda\kappa}^\alpha g^{\lambda\kappa}\tau^{\eta\beta})_{,\beta}\nonumber\\
&+&\frac{1}{3}\Psi_{\eta\gamma}^\iota g^{\eta\gamma}\tau^{\alpha\beta}\tau^{\lambda\kappa}g_{\lambda\kappa} g_{\alpha\iota ,\beta}\nonumber\\
\Gamma^\mu_{\alpha\beta}=&&
\begin{Bmatrix} 
\mu \\ \alpha\beta
\end{Bmatrix}
+\frac{1}{2\sigma}X^\mu X^\eta g_{\alpha\beta,\eta}\nonumber\\
&-&\frac{1}{6\sigma}\Bigl(
\delta^\mu_\alpha X^\nu X^\eta g_{\beta\nu,\eta}+\delta^\mu_\beta X^\nu X^\eta g_{\alpha\nu,\eta}\Bigr)\nonumber
\end{eqnarray}
\noindent It is useful to recall the well-known result from differential geometry; \textit{all two-dimensional manifolds are conformally flat}. Hence, the transport coefficients out of Onsager's region can always be brought into the form
\begin{equation}\label{2dim5}
g_{\mu\nu}=L_{\mu\nu}\exp{\phi (X)}
\end{equation}
\noindent with $\phi$ denoting a scalar field depending on the thermodynamic forces. By plugging Eq.~(\ref{2dim5}) into Eqs~(\ref{2dim3}) and (\ref{2dim4}) we get the PDE which has to be solved for the conformal field $\phi$. In this case, the Onsager condition requires $\phi(0)=0$.

\noindent Concerning the action, we adopt the expression proposed in literature \cite{jackiw2}. This action reads
\begin{equation}\label{2dim6}
I=\int {\mathcal N}(B+T){\sqrt g}\ d^2X
\end{equation}
\noindent where ${\mathcal N}$ is an auxiliary scalar field (analogous to the {\it dilaton field} \cite{cavaglia}), which plays the role of a Lagrangian multiplier. Notice that in this formalism, the dynamical fields present in the action~(\ref{2dim6}) are the dilatation field and the transport coefficients. In this case the affine connection does not play the role of an independent field (it is a dynamical variable only when $n >2$) and it intervenes in the dynamics through the second expression of Eqs~(\ref{2dim4}). By varying this action with respect to ${\mathcal N}$ we get Eq.~(\ref{2dim3}), while variation with respect to the transport coefficients yields the PDE for ${\mathcal N}$. The PDE for the transport coefficients is decoupled from that for the dilaton field. However, as we will see in the next work, this will not be the case when the skew-symmetric part of the transport coefficients is different from zero.

\noindent Let us now determine the nonlinear partial differential equation satisfied by a conformal factor $\Lambda$ of the metric $g_{\mu\nu}$. A conformal manifold is a manifold equipped with an equivalence class of metric tensors, in which two metrics $g_{\mu\nu}$ and ${\tilde g}_{\mu\nu}$ are equivalent if and only if
\begin{equation}\label{conf1}
g_{\mu\nu}=\Lambda(X) {\tilde g}_{\mu\nu}
\end{equation}
\noindent where $\Lambda(x)$ is a real-valued smooth function defined on the manifold referred to as \textit{conformal factor}. An equivalence class of such metrics is known as a conformal metric or conformal class. Thus, a conformal metric may be regarded as a metric that is only defined {\it up to scale}. A conformal metric is conformally flat if there is a metric representing it that is flat, i.e.
\begin{equation}\label{conf2}
g_{\mu\nu}=\Lambda(X) L_{\mu\nu}
\end{equation}
\noindent Often conformal metrics are treated by selecting a metric in the conformal class, and applying only {\it conformally invariant} constructions to the chosen metric. From Eq.~(\ref{conf2}) we get
\begin{eqnarray}\label{conf3}
&&g^{\mu\nu}\!=\!\frac{1}{\Lambda} L^{\mu\nu};\ g\!=\!\Lambda^nL;\ L_{\mu\nu}X^\nu\!=\!\Lambda^{-1}X_\mu;\ L^{\mu\nu}X_\nu=\Lambda x^\mu\nonumber\\
&&\Gamma^\lambda_{\mu\nu}=\frac{1}{2\Lambda}\left(\Lambda_{, \nu}\delta^\lambda_\mu+\Lambda_{, \mu}\delta^\lambda_\nu-\Lambda_{,\kappa}L^{\kappa\lambda}_{\mu\nu}\right)+\frac{X^\lambda}{2\sigma}L_{\mu\nu}X^\kappa\Lambda_{,\kappa}\nonumber\\
&&\qquad\ \ -\frac{\Lambda_{,\kappa} X^\kappa}{2(n+1)\Lambda\sigma}\left(X_\nu\delta^\lambda_\mu+X_\mu\delta^\lambda_\nu\right)
\end{eqnarray}
\noindent For the two-dimensional case, we have
\begin{eqnarray}\label{conf4}
T&=&\frac{2}{9\Lambda\sigma}\Lambda_{,\lambda}X^\lambda+\frac{2}{9\Lambda\sigma}\Lambda_{,\lambda ,\kappa}X^\lambda X^\kappa+\frac{4}{9\Lambda^2\sigma}\Lambda_{,\lambda}\Lambda_{,\kappa}X^\lambda X^\kappa\nonumber\\
B&=&\frac{\Lambda_{,\lambda ,\kappa}L^{\lambda\kappa}}{\Lambda^2}-\frac{\Lambda_{,\lambda}\Lambda_{,\kappa}L^{\lambda\kappa}}{\Lambda^3}-\frac{2\Lambda_{,\lambda}X^\lambda}{3\Lambda\sigma}-\frac{2\Lambda_{,\lambda ,\kappa}X^\lambda X^\kappa}{3\Lambda\sigma}\nonumber\\
&+&\frac{5\Lambda_{,\lambda}\Lambda_{, \kappa} X^\lambda X^\kappa}{9\Lambda^2\sigma}
\end{eqnarray}
\noindent where Eqs~(\ref{TAP3}), (\ref{nte6}) and (\ref{conf3}) have been taken into account. From Eq.~(\ref{2dim3}) we get 
\begin{eqnarray}\label{conf5}
B+T&=&\sigma_{(L)}\bigl(\Lambda\Lambda_{,\mu ,\nu}-\Lambda_{,\mu}\Lambda_{,\nu}\bigr)L^{\mu\nu}-\\
&&\Bigl(\frac{4}{9}\Lambda \Lambda_{,\mu ,\nu}-\Lambda_{, \mu} \Lambda_{,\nu}\Bigr)X^\mu X^\nu-\frac{4}{9}\Lambda \Lambda_{,\mu}X^\mu=0\nonumber
\end{eqnarray}
\noindent Now, by setting $\Lambda(X)=\exp (\phi(X))$, Eq.~(\ref{conf5}) reads \cite{sonnino8}
\begin{eqnarray}\label{conf6}
L^{\mu\nu}\frac{\partial^2\phi}{\partial X^\mu\partial X^\nu}&-&\frac{4}{9\sigma_{(L)}}X^\mu X^\nu\frac{\partial^2\phi}{\partial X^\mu\partial X^\nu}-\frac{4}{9\sigma_{(L)}}X^\mu\frac{\partial\phi}{\partial X^\mu}+\nonumber\\
&&\frac{5}{9\sigma_{(L)}}\left(X^\mu\frac{\partial\phi}{\partial X^\mu}\right)^2=0
\end{eqnarray}
\noindent Eq.~(\ref{conf6}) has to be solved for the conformal factor $\phi$. By introducing the differential operators, invariant under TCT \cite{sonnino1},
\begin{equation}\label{conf7}
{\mathcal O}\equiv X^\mu\frac{\partial}{\partial X^\mu},\quad \Box^2\equiv L^{\mu\nu}\frac{\partial^2}{\partial X^\mu\partial X^\nu}
\end{equation}
\noindent Eq.~(\ref{conf6}), can be cast into a manifestly TCT-covariant form 
\begin{equation}\label{conf8}
\left(9\sigma_{(L)}\Box^2-4{\mathcal O}^2\right)\phi+5\left({\mathcal O}\phi\right)^2=0
\end{equation}
\noindent Let us now performing the following coordinate transformation \footnote{Note that linear transformations of coordinates are allowed because this class of transformations belong to the TCT-group \cite{sonnino1}.}:
\begin{equation}\label{conf9}
{X'}^\lambda=A^\lambda_\kappa X^\kappa,\quad{\rm with}\ \ A^\mu_\nu\ {\rm such\ that}\ \ A^\alpha_\lambda L^{\lambda\kappa}A^\beta_\kappa={\rm I}^{\alpha\beta}
\end{equation}
\noindent with ${\rm I}^{\alpha\beta}$ denoting the Identity matrix. Notice that, since the matrix $L_{\mu\nu}$ is a positive definite matrix, there exists always a matrix $A^\mu_\nu$, which satisfies condition~(\ref{conf9}). Finally, we get 
\begin{align}\label{conf10}
&\left(9\sigma'_{(L)}\Box^{'2}-4{\mathcal O}^{'2}\right)\phi'+5\left({\mathcal O}'\phi'\right)^2=0\quad{\rm with}\\
&\sigma'_{(L)}={X^{'1}}^2+{X^{'2}}^2,\quad \Box^{'2}\equiv \frac{\partial^2}{\partial {X^{'1}}^2}+\frac{\partial^2}{\partial {X^{'2}}^2}\nonumber
\end{align}
\noindent Eq.~(\ref{conf8}) (or, equivalently, Eq.~(\ref{conf10})) is, in Thermodynamical Field Theory (TFT), analogous to Liouville's equation in Riemannian (or pseudo-Riemannian) geometry \cite{sonnino8}.

\section{Linearised Transport Equations}\label{LTE}

\noindent When the transport coefficients is close to Onsager's matrix, we may set
\begin{equation}\label{l1}
g_{\mu\nu}\simeq L_{\mu\nu}+h_{\mu\nu}
\end{equation}
\noindent with $h_{\mu\nu}$ considered as a small perturbation of the transport matrix coefficients. We also introduce a small parameter $\varepsilon$ of the order of $\sigma^{-1}$ (considered as a small quantity). The linearised Transport Equations are obtained by discarding systematically in the following calculations:

\noindent {\bf i}) All the terms of order $hh,\ hhh,\cdots$;

\noindent {\bf ii)} All the terms of order $\varepsilon^2$, of order $\sigma^{-2}$ or of higher order.

\subsection{Linearised Transport Equations for $n > 2$}

\noindent At the dominant order in $h_{\mu\nu}$ we get for $n > 2$
\begin{equation}\label{l2a}
 B_{\mu\nu}\!=\!T_{\mu\nu}\!-\!\frac{1}{2}L_{\mu\nu}\left(\Psi^\beta_{\lambda\kappa}L^{\lambda\kappa}\!+\!\frac{2}{(n+1)(n-2)}\Psi^\lambda_{\mu\nu}L^{\mu\nu}L_{\lambda\kappa}\tau_{(L)}^{\kappa\beta}\right)_{, \beta}\!=\!W^{(S)}_{\mu\nu}
\end{equation}
\noindent where
\begin{eqnarray}\label{l3}
B_{\mu\nu}&\simeq& \frac{\partial \Gamma^\lambda_{\lambda\mu}}{\partial X^\nu}-\frac{\partial\Gamma^\lambda_{\mu\nu}}{\partial X^\lambda}\qquad\quad {\rm hence}\\
B_{\mu\nu}&\simeq& \frac{1}{2}\Bigl(L^{\lambda\kappa}\frac{\partial^2 h_{\mu\nu}}{\partial X^\lambda X^\kappa}+L^{\lambda\kappa}\frac{\partial^2 h_{\lambda\kappa}}{\partial X^\mu X^\nu}-L^{\lambda\kappa}\frac{\partial^2 h_{\kappa\mu}}{\partial X^\lambda X^\nu}-L^{\lambda\kappa}\frac{\partial^2 h_{\kappa\nu}}{\partial X^\lambda X^\mu}\nonumber\\
&-&\frac{\partial}{\partial X^\lambda}\bigl(\tau^{\lambda\kappa}\frac{\partial h_{\mu\nu}}{\partial X^\kappa}\bigr)+\frac{1}{n+1}\frac{\partial}{\partial X^\mu}\bigl(\tau^{\lambda\kappa}\frac{\partial h_{\nu\kappa}}{\partial X^\lambda}\bigr)+\frac{1}{n+1}\frac{\partial}{\partial X^\nu}\bigl(\tau^{\lambda\kappa}\frac{\partial h_{\mu\kappa}}{\partial X^\lambda}\bigr)\Bigr)\nonumber
\end{eqnarray}
\noindent or
\begin{eqnarray}\label{l4}
B_{\mu\nu}&=&B^{(0)}_{\mu\nu}+B^{(1)}_{\mu\nu}\qquad {\rm with}\\
B^{(0)}_{\mu\nu}&=&\frac{1}{2}\Bigl(L^{\lambda\kappa}\frac{\partial^2 h_{\mu\nu}}{\partial X^\lambda X^\kappa}+L^{\lambda\kappa}\frac{\partial^2 h_{\lambda\kappa}}{\partial X^\mu X^\nu}-L^{\lambda\kappa}\frac{\partial^2 h_{\kappa\mu}}{\partial X^\lambda X^\nu}-L^{\lambda\kappa}\frac{\partial^2 h_{\kappa\nu}}{\partial X^\lambda X^\mu}\Bigr)\nonumber\\
B^{(1)}_{\mu\nu}&=&\frac{1}{2}\Bigl(-\frac{\partial}{\partial X^\lambda}\bigl(\tau^{\lambda\kappa}\frac{\partial h_{\mu\nu}}{\partial X^\kappa}\bigr)+\frac{1}{n+1}\frac{\partial}{\partial X^\mu}\bigl(\tau^{\lambda\kappa}\frac{\partial h_{\nu\kappa}}{\partial X^\lambda}\bigr)+\frac{1}{n+1}\frac{\partial}{\partial X^\nu}\bigl(\tau^{\lambda\kappa}\frac{\partial h_{\mu\kappa}}{\partial X^\lambda}\bigr)\Bigr)\nonumber
\end{eqnarray}
\noindent We also have for $n > 2$
\begin{eqnarray}\label{l5}
T_{\mu\nu}=&-&\Psi^\beta_{\mu\nu, \beta}+\frac{1}{2}\Psi^\beta_{\lambda\kappa ,\beta}L^{\lambda\kappa}L_{\mu\nu}-\frac{1}{2}\left((\Psi^\lambda_{\kappa\nu}L_{\lambda\mu}+\Psi^\lambda_{\kappa\mu}L_{\lambda\nu})\tau_{(L)}^{\kappa\beta}\right)_{, \beta}\\
&+&\frac{1}{2(n+1)}\left(\Psi^\lambda_{\eta\alpha}\tau^{\kappa\beta}_{(L)}+\Psi^\kappa_{\eta\alpha}\tau^{\lambda\beta}_{(L)}\right)_{, \beta}L^{\eta\alpha}L_{\lambda\mu}L_{\kappa\nu}\nonumber\\
\Psi^\mu_{\lambda\kappa}=&&\frac{1}{2}\tau_{(L)}^{\mu\eta}h_{\lambda\kappa,\eta}-\frac{1}{2(n+1)}\tau_{(L)}^{\eta\alpha}h_{\lambda\eta,\alpha}\delta^\mu_\kappa-\frac{1}{2(n+1)}\tau_{(L)}^{\eta\alpha}h_{\kappa\eta,\alpha}\delta^\mu_\lambda\nonumber\\
T=&&T_{\mu\nu}L^{\mu\nu}=\left(\frac{n-2}{2}\Psi^\beta_{\mu\nu}L^{\mu\nu}+\frac{1}{n+1}\Psi^\lambda_{\mu\nu}L^{\mu\nu}L_{\lambda\kappa}\tau_{(L)}^{\kappa\beta}\right)_{, \beta}\nonumber\\
B=&-&\left(\Psi^\beta_{\mu\nu}L^{\mu\nu}+\frac{2}{(n+1)(n-2)}\Psi^\lambda_{\mu\nu}L^{\mu\nu}L_{\lambda\kappa}\tau_{(L)}^{\kappa\beta}\right)_{, \beta}\nonumber
\end{eqnarray}
\noindent with
\begin{equation}\label{l6}
\tau_{(L)}^{\mu\nu}=\frac{X^\mu X^\nu}{\sigma_{(L)}}\quad ;\quad\sigma_{(L)}=L_{\mu\nu}X^\mu X^\nu
\end{equation}
\noindent After a little algebra, we find that identity~(\ref{nte9}) (or Eq.~(\ref{nte8})) implies
\begin{equation}\label{l7}
\frac{\partial T^{\mu\nu}}{\partial X^\nu}=0
\end{equation}
\noindent This {\it conservation law} is consistent with the fact that our PDE (and the Lagrangian) are invariant under the TCT. Hence, for the Noether theorem, to this invariance is associated a {\it conserved current} and, so, a conserved {\it source tensor} \cite{sonnino6}.

\noindent As mentioned above, one way to get the approximate solution of the Eq.~(\ref{l2a}) is to introduce a parameter $\varepsilon$ of the order of $\sigma^{-1}$, which we consider to be a small quantity. By setting 
\begin{equation}\label{ap1}
h_{\mu\nu}\simeq h_{\mu\nu}^{(0)}+\varepsilon h_{\mu\nu}^{(1)}\quad {\rm with}\quad \varepsilon\sim O(\sigma^{-1})
\end{equation}
\noindent the linearised Transport Equations for $n > 2$ read
\begin{eqnarray}\label{I8}
&&B_{\mu\nu}^{(0)}(h^{(0)})=0\\
&&B_{\mu\nu}^{(0)}(\epsilon h^{(1)})=T_{\mu\nu}(h^{(0)})-B^{(1)}_{\mu\nu}(h^{(0)})-\frac{1}{2}L_{\mu\nu}\left(\bigl(\Psi^\beta_{\lambda\kappa}(h^{(0)})L^{\lambda\kappa}\right)_{, \beta}\nonumber\\
&&\qquad\qquad\quad -\frac{1}{(n+1)(n-2)}L_{\mu\nu}\left(\Psi^\alpha_{\lambda\kappa}(h^{(0)})L^{\lambda\kappa}L_{\alpha\eta}\tau^{\eta\beta}_{(L)}\right)_{,\beta}\equiv W_{\mu\nu}^{(S)}(h^{(0)})\nonumber
\end{eqnarray} 
\noindent Note that $W^{(S)}_{\mu\nu}(h^{(0)})\rightarrow 0$ as $h^{(0)}\rightarrow 0$. 

\subsection{Examples of Simplification of the Linearised Transport Equations}\label{simplyeq}

\noindent It is worth mentioning that in several cases the second PDE of system~(\ref{I8}) simplifies significantly. Indeed, we have already noticed that tensor $W^{(S)}_{\mu\nu}$ satisfies the identity
\begin{equation}\label{add1}
L^{\mu\lambda}\frac{\partial W^{(S)}_{\lambda\nu}}{\partial X^\mu}-\frac{1}{2}L^{\mu\lambda}\frac{\partial W^{(S)}_{\mu\lambda}}{\partial X^\nu}\equiv 0
\end{equation}
\noindent Now, let us suppose to have solved the following Poisson PDE
\begin{equation}\label{add2}
L^{\lambda\kappa}\frac{\partial ^2 {h}_{\mu\nu}^{(1)}}{\partial X^\lambda\partial X^\kappa}=W^{(S)}_{\mu\nu}\qquad {\rm with}\quad {h}^{(1)}_{\mu\nu}\mid_{\partial\Omega}=0
\end{equation}
\noindent with $\partial\Omega$ denoting the boundary\footnote{Note that the boundary conditions have already been satisfied at zero order, and for this reasons, ${\bar h}^{(1)}_{\mu\nu}$ should vanish at the boundary.}. From Eqs~(\ref{add1}) and (\ref{add2}), we get
\begin{equation}\label{add3}
L^{\lambda\kappa}\frac{\partial ^2 {\mathcal R}_\nu(x)}{\partial X^\lambda\partial X^\kappa}=0\qquad {\rm with}\quad {\mathcal R}_\nu(x)\equiv \frac{1}{2}L^{\mu\lambda}\frac{{\partial}{h}_{\mu\lambda}}{\partial X^{\nu}}-L^{\mu\lambda}\frac{{\partial}{h}_{\lambda\nu}}{\partial X^{\mu}}
\end{equation}
\noindent Hence, if it happens, for example, that  ${\mathcal R}_\nu(x)\mid_{\partial\Omega}=0+\mathcal{O}(\varepsilon^2)$ we also have ${\mathcal R}_\nu(x)=0+\mathcal{O}(\varepsilon^2)$ throughout the space. In other words, if it happens, for example, that the derivative of the perturbation $h_{\mu\nu}^{(1)}$ vanishes at the boundary
\begin{equation}\label{add4}
\frac{\partial h_{\mu\nu}^{(1)}}{\partial X^{\lambda}}{\Bigg\arrowvert}_{\partial\Omega}=0
\end{equation}
\noindent the second PDE of system~(\ref{I8}) reduces to a Poisson's PDE and the Transport Equations to be solved reduce to
\begin{eqnarray}\label{add5}
&&B_{\mu\nu}^{(0)}(h^{(0)})=\ 0 \\
&&L^{\lambda\kappa}\frac{\partial^2 {\varepsilon h_{\mu\nu}^{(1)}}(X)}{\partial X^{\lambda} \partial {X^\kappa}}=W_{\mu\nu}^{(S)}(X)\nonumber
\end{eqnarray} 
\noindent There is another important case where the second PDE of system~(\ref{I8}) reduces to a Poisson PDE. This happens when the perturbation takes the form 
\begin{equation}\label{add6}
h_{\mu\nu}(X)=L_{\mu\nu} h(X)
\end{equation}
\noindent with $h(X)$ indicating a scalar field. As we shall see in Section~\ref{linsol1}, this is exactly what happens for the two-dimensional case (see Eq.~(\ref{2dim5})). PDEs~(\ref{add5}) should be solved with the boundary conditions specified in the Annex.
\vskip 0truecm
\subsection{Linearised Transport Equation for $n=2$}\label{twodim}
\noindent As seen in Section~\ref{2dim}, for $n=2$ the PDE to be solved is
\begin{equation}\label{n2}
B=-T
\end{equation}
\noindent Hence, the linearised Transport Equation read
\begin{eqnarray}\label{n21}
B&=&0\qquad\quad {\rm for\ the\ homogeneous\ case}\\
B&=&-T_{(L)}\quad{\rm for\ the\ inhomogeneous\ case}\nonumber
\end{eqnarray}
\noindent with
\begin{equation}\label{n22}
T_{(L)}=\frac{1}{3}L_{\eta\alpha}L^{\lambda\kappa}\left(\Psi^\alpha_{\lambda\kappa}\tau_{(L)}^{\eta\beta}\right)_{, \beta}
\end{equation}
\noindent We have already mentioned that for $n=2$ the solution of Eqs~(\ref{n21}) can always be brought into the form (see Eq.~(\ref{2dim5}) in Section~\ref{2dim})
\begin{equation}\label{n23}
g_{\mu\nu}=L_{\mu\nu}f(x)
\end{equation}
\noindent By setting 
\begin{equation}\label{n25}
h_{\mu\nu}\simeq h_{\mu\nu}^{(0)}+\varepsilon h_{\mu\nu}^{(1)}\quad {\rm with}\quad \varepsilon\sim O(\sigma^{-1})
\end{equation}
\noindent we get the linearised Transport Equations for $n=2$
\begin{eqnarray}\label{n26}
&&B^{(0)}(h^{(0)})=B^{(0)}_{\mu\nu}L^{\mu\nu}(h^{(0)})=0\\
&&B^{(0)}(\epsilon h^{(1)})=B^{(0)}_{\mu\nu}L^{\mu\nu}(\epsilon h^{(1)})=-B^{(1)}_{\mu\nu}L^{\mu\nu}(h^{(0)})-T_{(L)}(h^{(0)})=W^{(S)}(x)\nonumber
\end{eqnarray}
\noindent The analytic solution of system~(\ref{n26}) (or system~(\ref{n21})) can be found in the Section~\ref{linsol1}.

\section{TFT Gauge Invariance}\label{TFTGI}
In field theories, different configurations of the unobservable fields can result in identical observable quantities. A transformation from one such field configuration to another is called a \textit{gauge transformation}; the lack of change in the measurable quantities, despite the field being transformed, is a property called \textit{gauge invariance}. In this Section we shall clarify the physical meaning of the gauge invariance in the Thermodynamical Field Theory. To carry out this task we need first to recall some fundamental theorems concerning the solution of the differential equations~(\ref{I8}). After this, in subsection~(\ref{GI}) we provide the physical interpretation of the gauge invariance in the TFT.

\subsection{Basic Theorems for the PDEs $B^{(0)}_{\mu\nu}(h)=W^{(S)}_{\mu\nu}$}\label{theo}

\noindent Let us consider the PDE $B_{\mu\nu}^{(0)}(h)=W^{(S)}_{\mu\nu}$ where the source $W^{(S)}_{\mu\nu}$ may be either different from zero or absent. By direct inspection, we find that if $h_{\mu\nu}(X)$ is a solution of $B_{\mu\nu}^{(0)}(h)=W^{(S)}_{\mu\nu}$, then so will be
\begin{equation}\label{B3}
{\widehat h}_{\mu\nu}(X)=h_{\mu\nu}(X)+\frac{\partial u_\nu(X)}{\partial X^\mu}+ \frac{\partial u_\mu (X)}{\partial X^\nu}
\end{equation} 
\noindent where $u_\mu(X)$ are $n$ small but otherwise arbitrary functions of $X^\mu$. Hence, tensor $B_{\mu\nu}^{(0)}(h)$ is unaffected by gauge transformations~(\ref{B3}). Thanks to this gauge-invariance, we have the following theorem \cite{wheeler}: 
\begin{theorem}
If one knows a specific solution ${\bar h}_{\mu\nu}$ to the linearised equations (\ref{I8}) for a given $T_{\mu\nu}$ one can obtain another solution that describes precisely the same physical situation by the change of gauge (\ref{B3}), in which $u_\mu$ are  arbitrary but small functions.
\end{theorem}
\noindent So, if we are able to find a particular solution of Eqs~(\ref{I8}), say ${\bar h}_{\mu\nu}(X)$, all the other solutions ${\widehat h}_{\mu\nu}(X)$ can be found by adding to the particular solution ${\bar h}_{\mu\nu}(X)$ the tensor $\frac{\partial u_\nu(X)}{\partial X^\mu}+ \frac{\partial u_\mu(X)}{\partial X^\nu}$. In addition, ${\widehat h}_{\mu\nu}(X)$ and $h_{\mu\nu}(X)$ possess the {\it same physical meaning}.

\noindent We also have the following
\begin{theorem}
If one knows a specific solution ${\bar h}_{\mu\nu}$ to the second equation of system~(\ref{I8}) for a given $W_{\mu\nu}$, it is always possible to choose $u_\mu$ such that the new solution ${\widehat h}_{\mu\nu}(X)$ satisfies the gauge
\begin{eqnarray}\label{B4}
\frac{1}{2}L^{\lambda\kappa}\frac{{\partial}{\widehat h}_{\lambda\kappa}}{\partial X^{\nu}}&=&L^{\lambda\kappa}\frac{{\partial}{\widehat h}_{\lambda\nu}}{\partial X^{\kappa}}\qquad {\rm with}\\
{\widehat h}_{\mu\nu}(X)&=&{\bar h}_{\mu\nu}(X)+\frac{\partial u_\nu(X)}{\partial X^\mu}+ \frac{\partial u_\mu(X)}{\partial X^\nu}\nonumber
\end{eqnarray} 
\end{theorem}
\noindent Indeed, ${\widehat h}_{\mu\nu}(X)$ manifestly satisfies the second equation of system~(\ref{I8}), and the gauge-condition [i.e. the first equation of Eqs~(\ref{B4})] is satisfied by choosing $u_\nu$ such that
\begin{equation}\label{B5}
L^{\lambda\kappa}\frac{\partial^2u_\nu(X)}{\partial X^{\lambda} \partial {X^\kappa}}=\frac{1}{2}L^{\lambda\kappa}\frac{{\bar h}_{\lambda\kappa}(X)}{\partial X^\nu}-L^{\lambda\kappa}\frac{{\bar h}_{\kappa\nu }(X)}{\partial X^\lambda}
\end{equation} 
\noindent Note that, thanks to Eq.~(\ref{B5}), ${\widehat h}_{\mu\nu}(X)$ satisfies the second PDE of system~(\ref{I8}) because it satisfies {\it simultaneously} the gauge-condition (\ref{B4}) and the following Poisson's PDE
\begin{equation}\label{B6}
L^{\lambda\kappa}\frac{\partial^2 {\widehat h}_{\mu\nu}(X)}{\partial X^{\lambda} \partial X^{\kappa}}=W^{(S)}_{\mu\nu}(X)
\end{equation} 
\noindent In conclusion, if we know a specific solution ${\bar h}_{\mu\nu}(X)$, thanks to Eqs~(\ref{B4})-(\ref{B5}), we shall also able to get the expression for ${\widehat h}_{\mu\nu}(X)$ satisfying simultaneously the gauge-condition and Poisson's PDE~(\ref{B6}).

\noindent Incidentally, we also have the following theorem \cite{wheeler}:

\begin{theorem}
By performing the following change of variables $X^\mu\rightarrow X^{'\mu}$ 
\begin{equation}\label{B7}
X^{'\mu}=X^\mu+L^{\mu\nu} u_\nu(X)
\end{equation}
\end{theorem}
\noindent {\it where} $u_\nu$ {\it is the solution} of Eq.~(\ref{B5}), {\it the transformed tensor of the unknown} $h_{\mu\nu}(X)$ {\it is a solution of Poisson's PDE}~(\ref{B6}).

\noindent Indeed, by direct inspection, we can check that we find exactly the same PDE for the transformed tensor $h'_{\mu\nu}(X')$, obtained by $h_{\mu\nu}$ after the coordinate transformation $X^\mu\rightarrow X^{'\mu}$. The only precaution to be taken is to remain within the limits of validity of the {\it weak-field approximation} and, therefore, non-linear terms of the type $hh$, $hu$, $uu$, and of higher order must be neglected.

\subsection{A Note on the Physical Meaning of the Gauge Invariance}\label{GI}
Let us consider a group of transformations of the field variables that leaves unchanged the basic physical observable. This group of transformations is called \textit{gauge-transformations} and a theory where \textit{all} the basic observables are unchanged under a gauge transformation of the field variables is referred to as a \textit{gauge-invariant theory}. In Electrodynamics, for example, the Lorentz transformation or the Coulomb transformation are both gauge-transformations since they do not affect the values of the electrodynamic observables, i.e. the values of the electric and the magnetic fields. In other words, the experimentalist is not able to detect the gauge-transformation choice, with any kind of system, and he is unable to notice any difference between two different gauge-choices.

\noindent Another example is the theory of the General Relativity (GRT) since transformations~(\ref{B3}) leave unchanged the physical observables i.e., the Ricci tensor, the Einstein tensor and the Riemannian curvature tensor. Hence, in GRT, transformations~(\ref{B3}) may be regarded as the gauge-transformations. 

\noindent Thus, \textit{the concept of gauge invariance is intimately related to the one of unchanged physical observables}. This means that, before starting calculations, we should firstly identify all the physical observables linked to these transformations and, successively, check whether the values of these observables may be affected by a field-variables transformation. 

\noindent The following example will made clear the concept. Let us suppose  (absurdly, of course) that, in classical Electrodynamics, the physical observables are not only the electric field {\bf E} and the magnetic field {\bf B}, but also the scalar potential $\phi$ and the vector potential {\bf A}. Thus, we suppose that an experimentalist is able to measure, with its instruments, also the numerical values of these two variables (in addition to the electromagnetic fields {\bf E} and {\bf B}). The electromagnetic fields {\bf E} and {\bf B} will still remain unaffected under Lorentz's or Coulomb's transformations. However, the question is: \textit{May we still consider this New Electrodynamics as a gauge-invariant theory ?} The answer is No. The only thing that has been changed is the fact that, in this \textit{new Electrodynamics}, the experimentalist is now able to measure the scalar potential $\phi $ and the potential vector {\bf A} (in addition to the electric and the magnetic fields). What happens then ? If we perform calculations by using the Lorentz transformation, as to the electric field and the magnetic field, the experimentalist will confirm the good agreement between the theoretical predictions and the experimental data. However, in general, he will find a discrepancy between the experimental data and the values of the scalar potential ${\phi}_L$ and of the potential vector ${\bf A}_L$ established by the Lorentz transformation. For the experimentalist, the only way to overcome this impasse is to know the mathematical expression linking the scalar potential ${\phi}_L$ and the potential vector ${\bf A}_L$ with the scalar potential ${\phi}_{Exp.}$ and the potential vector ${\bf A}_{Exp.}$ measured in laboratory. In this case, the \textit{new Electrodynamics} looses its status of \textit{gauge-invariant theory}.

\noindent Let us now consider another comparison. Let us compare the Thermodynamical Field Theory (TFT) with the General Relativity Theory (GRT). Here, there is a clear divergence between the TFT and the GRT. Indeed, as mentioned above, the physical observables in the GRT are the Ricci tensor, the Einstein tensor and the Riemannian curvature tensor. Transformations~(\ref{B3}) leave unchanged these physical observables. Hence in GRT, \textit{without loss of generality}, we may suppose that there exists a specific solution of the PDE $B_{\mu\nu}^{(0)}(h^{(0)})=0$. Note that it does matter if, in reality, we do not know the mathematical expression of this solution; the proof of its existence is just sufficient. For {\bf Theorem 2}, we may imagine to perform transformation~(\ref{B3}), with $u_\nu$ satisfying Eq.~(\ref{B5}), such that the new unknown reduces to the PDE~(\ref{B6}) (and it satisfies, at the same time, the gauge-condition). All of this is consistent with the General Covariance Principle (GCP), which allows choosing the coordinate system as we like such as, for example, the coordinate transformation~(\ref{B7}) with $u_\nu$ satisfying the PDE~(\ref{B5}), where the second PDE of system~(\ref{I8}) reduces to Eq.~(\ref{B6}) (ref. to {\bf Theorem 3} and \cite{wheeler}, \cite{weinberg}). In conclusion, in GRT we do not need to know the mathematical expression of a specific solution of the second equations of system~(\ref{I8}), and we may start calculations by solving directly Eq.~(\ref{B6}).

\noindent The case of TFT is utterly different. Firstly, we cannot evoke the validity of the General Covariance principle and, even more importantly, the physical observables are the unknown $h_{\mu\nu}$ (indeed, the $h_{\mu\nu}$ are the transport coefficients). Hence, in analogy with what we said concerning the case of the \textit{new Electrodynamics}, if we want to communicate with experimentalists we are compelled to find, firstly, a specific solution of Eqs~(\ref{I8}) (either analytically or numerically) and, successively, obtain the general solution by applying Eq.~(\ref{B3}). In conclusion the TFT does not possess the status of {\it gauge-invariant theory} even though the {\bf Theorem 1} and {\bf Theorem 2}, remain valid and they are very useful for performing calculations. 

\section {Solution of the Linearised Equations}\label{linsol1}

\noindent  As seen in subsection~\ref{theo}, to get a concrete expression of a solution of Eqs~(\ref{I8}), firstly we have to be able to find a specific solution of this PDE. Successively, according to {\bf Theorem 1}, all the other solutions can be obtained by Eq.~(\ref{B3}). {\bf Theorem 2} allows getting the solution satisfying Eq.~(\ref{B6}) by solving Eq.~(\ref{B5}).
\vskip0.2truecm
\noindent {\bf $\bullet$ Solution of the Transport Equations for the Two - Dimensional Case}

\noindent For the two-dimensional case, the PDE to be solved read
\begin{eqnarray}\label{new1}
B^{(0)}(h^{(0)})&=&0\\\
B^{(0)}(\epsilon h^{(1)})&=&-B^{(1)}(h^{(0)})-T_{(L)}(h^{(0)})=W^{(S)}(h^{(0)})\nonumber
\end{eqnarray}
\noindent We start by solving the homogeneous differential equation $B^{(0)}(h^{(0)})=0$. Since our task is to find a specific solution, we look for a solution of the form~(\ref{n23}) \cite{sonnino}. 
\begin{equation}\label{new2}
{\bar h}^{(0)}_{\mu\nu}(x)=L_{\mu\nu}{\bar h}^{0)}(x)
\end{equation}
\noindent We have
\begin{eqnarray}\label{new3}
B^{(0)}({\bar h}^{(0)})&=&\frac{1}{2}L^{\mu\nu}\biggl(
L_{\mu\nu}L^{\lambda\kappa}\frac{\partial^2 {\bar h}^{(0)}}{\partial X^\lambda X^\kappa}\\
&+&L^{\lambda\kappa}L_{\lambda\kappa}\frac{\partial^2 {\bar h}^{(0)}}{\partial X^\mu X^\nu}-L^{\lambda\kappa}L_{\kappa\mu}\frac{\partial^2 {\bar h}^{0)}}{\partial X^\lambda X^\nu}-L^{\lambda\kappa}L_{\kappa\nu}\frac{\partial^2 {\bar h}^{(0)}}{\partial X^\lambda X^\mu}\biggr)=0\nonumber
\end{eqnarray}
\noindent By noticing that the sum of the last three contributions on the r.h.s. of Eq.~(\ref{new3}) vanishes identically, we finally get
\begin{equation}\label{new4}
B^{(0)}({\bar h}^{(0)})=
L^{\lambda\kappa}\frac{\partial^2 {\bar h}^{(0)}}{\partial X^\lambda X^\kappa}=0
\end{equation}
\noindent The second PDE of system~(\ref{new1}) reduces to a Poisson PDE. Indeed, also in this case, we look for a special solution of the form 
\begin{equation}\label{new5}
{\bar h}_{\mu\nu}^{(1)}(X)=L_{\mu\nu}{\bar h}^{(1)}(X)
\end{equation}
\noindent By inserting Eq.~(\ref{new5}) to the second PDE of system~(\ref{new1}) we get
\begin{equation}\label{new6}
L^{\lambda\kappa}\frac{\partial ^2\epsilon {\bar h}^{(1)}}{\partial x^\lambda\partial x^\kappa}=\left(\frac{2}{3}\tau_{(L)}^{\alpha\beta}h^{(0)}_{, \alpha}-\frac{1}{6}L_{\eta\alpha}L^{\iota\varrho}\Psi^{\alpha}_{\iota\varrho}\tau_{(L)}^{\eta\beta}\right)_{, \beta}=W^{(S)}(h^{(0)})
\end{equation}
\noindent since the contribution 
\begin{equation}\label{new7}
L^{\lambda\kappa}L_{\lambda\kappa}\frac{\partial^2 \epsilon{\bar h}^{(1)}}{\partial X^\mu X^\nu}-L^{\lambda\kappa}L_{\kappa\mu}\frac{\partial^2 \epsilon{\bar h}^{1)}}{\partial X^\lambda X^\nu}-L^{\lambda\kappa}L_{\kappa\nu}\frac{\partial^2 \epsilon{\bar h}^{(1)}}{\partial X^\lambda X^\mu}\equiv 0
\end{equation}
\noindent vanishes identically. In conclusion, for $n=2$ the PDEs to be solved are
\begin{eqnarray}\label{new8}
&&L^{\lambda\kappa}\frac{\partial^2{\bar h}^{(0)}(X)}{\partial X^{\lambda}\partial X^{\kappa}}=0\\
&&L^{\lambda\kappa}\frac{\partial ^2\epsilon {\bar h}^{(1)}(X)}{\partial X^{\lambda}\partial X^{\kappa}}=\frac{1}{6} \left(4\tau_{(L)}^{\alpha\beta}{\bar h}^{(0)}_{, \alpha}-L_{\eta\alpha}L^{\iota\varrho}\Psi^{\alpha}_{\iota\varrho}\tau_{(L)}^{\eta\beta}\right)_{, \beta}=W^{(S)}(X)\nonumber
\end{eqnarray}
\noindent By performing the following orthogonal coordinate transformation
\begin{equation}\label{new9}
{X'}^\lambda=A^\lambda_\kappa X^\kappa
\end{equation}
\noindent Eqs.~(\ref{new8}) reads 
\begin{eqnarray}\label{new10}
A^\alpha_\lambda L^{\lambda\kappa}A^\beta_\kappa\frac{\partial^2{\bar h}^{'(0)}(X')}{\partial X^{'\alpha}\partial X^{'\beta}}&=&0\\
A^\alpha_\lambda L^{\lambda\kappa}A^\beta_\kappa\frac{\partial^2 \epsilon {\bar h}^{'(1)}(X')}{\partial X^{'\alpha}\partial X^{'\beta}}&=&\left(\frac{2}{3}\tau_{(L)}^{'\alpha\beta}{\bar h}^{'(0)}_{, \alpha}-\frac{1}{6}L_{\eta\alpha}L^{\iota\varrho}\Psi^{'\alpha}_{\iota\varrho}\tau_{(L)}^{'\eta\beta}\right)_{, \beta}\nonumber
\end{eqnarray}
\noindent Note that linear transformations of coordinates are allowed because this class of transformations belong to the group TCT \cite{sonnino1}. Since the tensor $L^{\mu\nu}$ is symmetric positive definite matrix, it is always possible to determine $A^\lambda_\kappa$ such that 
\begin{equation}\label{new11}
A^\alpha_\lambda L^{\lambda\kappa}A^\beta_\kappa={\rm I}^{\alpha\beta}
\end{equation}
\noindent with ${\rm I}^{\alpha\beta}$ denoting the components of the identity matrix. As a consequence, Eq.~(\ref{new10}) reads
\begin{eqnarray}\label{new12}
{\rm I}^{\alpha\beta}\frac{\partial^2{\bar h}^{'(0)}(X')}{\partial X^{'\alpha}\partial X^{'\beta}}&=&0\\
{\rm I}^{\alpha\beta}\frac{\partial ^2 \epsilon {\bar h}^{'(1)}(X')}{\partial x^{'\alpha}\partial X^{'\beta}}&=& \left(\frac{2}{3}\tau_{(L)}^{'\alpha\beta}{\bar h}^{'(0)}_{, \alpha}-\frac{1}{6}L_{\eta\alpha}L^{\iota\varrho}\Psi^{'\alpha}_{\iota\varrho}\tau_{(L)}^{'\eta\beta}\right)_{, \beta}=W^{'(S)}(X')\nonumber
\end{eqnarray}
\noindent Tensor ${\bar h}^{(i)}_{\mu\nu}(x)$ (with $i=0,1$) are obtained by tensors ${\bar h}^{'(i)}_{\mu\nu}(X')$ by means of the coordinate transformation~(\ref{new9}), with $A^\mu_\nu$ determined by Eq.~(\ref{new11}). We conclude this part of our analysis by noticing that the conformal field $\phi$, defined in Eq.~(\ref{2dim5}), is determined by ${\bar h}^{(i)}$ (with $i=0,1$) through the expression
\begin{equation}\label{new13}
\phi=\ln\bigl({1+{\bar h}^{0)}+(\epsilon {\bar h}^{(1)})\cdots}\bigr)\simeq {\bar h}^{0)}+(\epsilon {\bar h}^{(1)})
\end{equation}
\noindent since ${\bar h}^{(i)}$ are non-negative scalar fields. Note that, by setting
\begin{equation}\label{new14}
\phi\simeq\phi_0+\varepsilon\phi_1,\quad{\rm with}\quad\varepsilon\sim O(\sigma_{(L)}^{-1})
\end{equation}
\noindent we get
\begin{eqnarray}\label{new15}
\Box^2\phi_0&=&0\\
\Box^2(\varepsilon\phi_1)&=&\frac{4}{9\sigma_{(L)}}{\mathcal O}^2\phi_0
\end{eqnarray}
\noindent which can be derived also by Eqs~(\ref{new12}). Eqs~(\ref{new12}), subject to the appropriate boundary conditions, admit solutions that can be obtained analytically. Let us consider the first equation of system~(\ref{new12}). This is a Laplacian PDE. According to the arguments illustrated in the Appendix, we have to solve the Laplacian equations in the first quadrant by imposing that the solutions vanish on the axes and they are constant on the arc of the circle of radius $R_0$. Successively, the entire solution, valid for all quadrants, is obtained by applying, the Schwartz principle \cite{CourantHilbert1}. In Appendix we report the solution of the Laplacian equation subjected to appropriate boundary conditions, for the case of two independent thermodynamic forces. We have \cite{sonnino}
\begin{equation}\label{an34}
{\bar h}^{'(0)}_{\mu\nu}(X^{'1}, X^{'2}) = \frac {2}{\pi}L_{\mu\nu}\arctan \biggl [\frac{4R_{0}^2 \mid X^{'1} X^{'2}\mid}{R_{0}^4-({X^{'1}}^{2}+{X^{'2}}^2)^2}\biggr ]
\end{equation}
\noindent where the new variables $\{X^{'\mu}\}$ (with $\mu=1,2$) are linked to the old ones $\{X^{\mu}\}$ by the (constant) $2\times 2$ matrix $A^\lambda_\eta$, which satisfies the relation
\begin{equation}\label{an35}
X^{'\mu}=A^\mu_\nu X^\nu\qquad{\rm with}\qquad A^\alpha_\lambda L^{\lambda\kappa}A^\beta_\kappa = {\rm I}^{\alpha\beta}
\end{equation}
\noindent The value of constant $\chi$ and the expression of the radius $R_0$ are determined by the thermodynamic system under consideration and by the specific problem to be solved. An example of calculation can be found in \cite{sonnino8}. It is worth mentioning that the method illustrated in Appendix applies also for obtaining the solution for $n$ independent thermodynamic forces. 

\noindent Let us now find the solution of the inhomogeneous problem. In this case, the second PDE of system~(\ref{I8}) reduces to a Poisson's PDE. Appendix reports the analytic solution of Poisson's PDE for a two-dimensional thermodynamic space. In polar coordinates $\rho,\theta$, we have
\begin{eqnarray}\label{an36}
\epsilon {\bar h}_{\mu\nu}^{'(1)}(\rho,\theta)=&-&L_{\mu\nu}R_0^2\sum_{n=1}^{\infty}\Bigl[\frac{\sin 2(n-1)\theta}{4(n-1)}\Bigl(\rho^{2(n-1)}\Bigl(\int_0^\rho t^{-2n+3}{\widehat W}^{'(S)}_n\!(t)dt-{\widehat \alpha}_n\Bigr)\\
& -&\rho^{-2(n-1)}\int_0^\rho t^{2n-1}{\widehat W}^{'(S)}_n\!(t)dt\Bigr)\Bigr]\nonumber
\end{eqnarray}
\noindent with
\begin{eqnarray}\label{an37}
{\widehat \alpha}_n&=&\int_0^1\left(t^{-2n+3}-t^{2n-1}\right){\widehat W}^{'(S)}_n\!(t)dt\qquad\quad {\rm and}\nonumber\\
{\widehat W}^{'(S)}_n\!(\rho)&\equiv&\frac{1}{\pi}\int_{-\pi}^{\pi}W^{'(S)}(\rho,\theta)\sin\left(2(n-1)\theta\right)d\theta\nonumber
\end{eqnarray}
\noindent where $\rho$ and $\theta$ are linked to the new variables $\{X^{'\mu}\}$ by the usual relations $\rho=({X^{'1}}^2+{X^{'2}}^2)^{1/2}$ and $\theta=\arctan(X^{'2}/X^{'1})$. Fig.~(\ref{Poisson1}) illustrates solution ${\bar h}^{'(1)}_{11}(\rho,\theta)$ in polar coordinates $\rho$ and $\theta$ in case of $W{'(S)}(X')\sim \varepsilon {\bar h}^{'(0)}(x')$ and $\chi=1$.
\begin{figure}[b]
\sidecaption
\includegraphics[width=4.0cm,height=4.0cm]{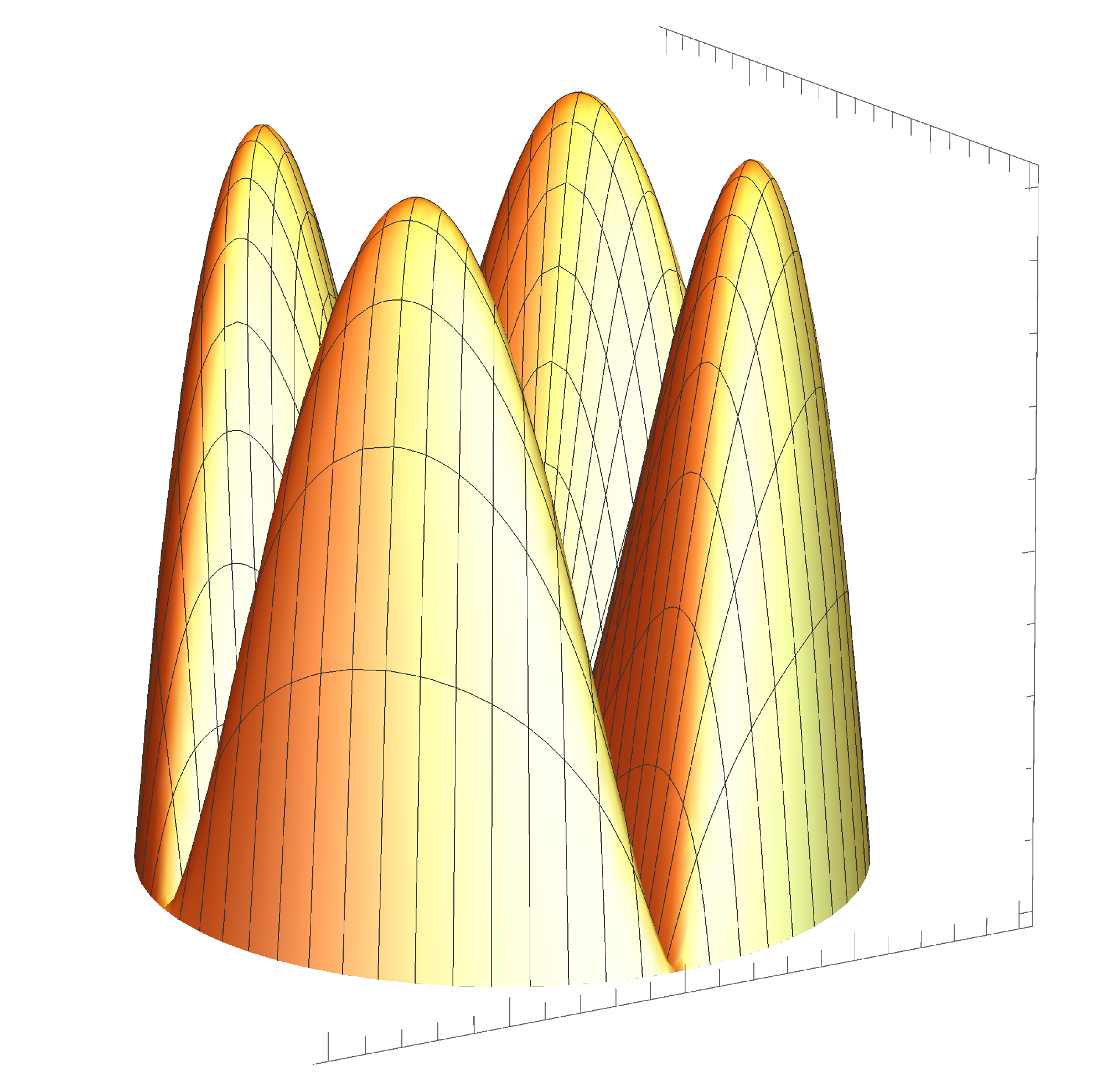}
\caption{ \label{Poisson1} ${\bar h}^{'(1)}_{11}(\rho,\theta)$ in polar coordinates $\rho$ and $\theta$ in case of $W^{'(S)}(X')\sim \varepsilon {\bar h}^{'(0)}(X')$ and $\chi=1$. }
\end{figure}

\section{Testing the validity of the PDE~(\ref{conf10}) - Computation of heat loss in L-mode, collisional FTU-plasma}\label{solution}
The aim of this Section is to test the validity of the PDE~(\ref{conf10}). To this purpose, we compare the theoretical predictions with the experimental data provided by the EUROfusion Consortium in Frascati (Rome-Italy) for FTU-plasmas \footnote{Here, we shall not enter in describing technical details \cite{sonnino9}, since all this is out of scope of the present work. Our aim is to show the good agreement between the proposed approach and experiments. The interested reader can be find a detailed description on the comparison between theory and the experimental data in FTU-plasmas in our article submitted for publication in a review specialised in the field of thermonuclear fusion.}. We started by comparing the theoretical predictions of Eq.~(\ref{conf10}) subjected to the correct boundary conditions, with the experimental data for FTU-plasmas, in a fully collisional regime. So, in a first phase, experiments have been performed in a zone of the Tokamak where the turbulent effects are almost {\it frozen}. In our calculations we have also taken into account the Shafranov-shift (which is not negligible in FTU-plasmas). The physical explanation of the Shafranov-shift is briefly sketched in Fig.~(\ref{fig_Shafranov_shift}).
\begin{figure}[b]
\sidecaption
\includegraphics[scale=.65]{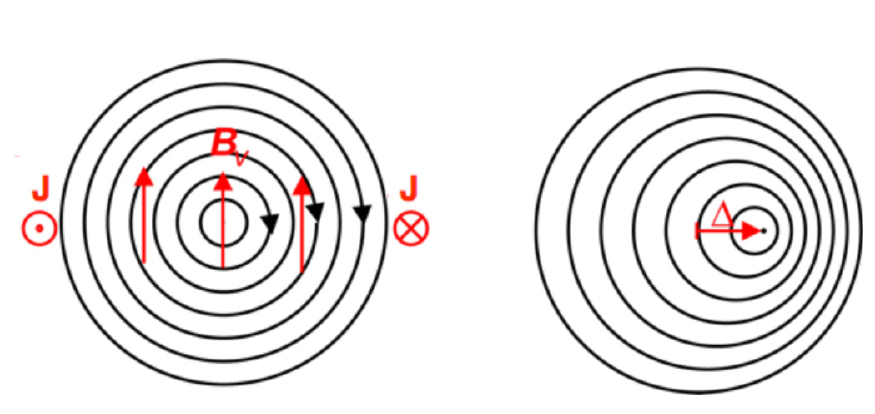}
\caption{{\bf The Shafranoshift shft}. In tokamak-plasmas, the plasma pressure leads to an outward shift $\Delta$ of the centre of the magnetic flux surfaces. ${\bf J}$ indicates the direction of the electric current that flows inside the plasma. Note that the poloidal magnetic field increases and the magnetic pressure can, then, balance the outward force \cite{sonnino6}.}
\label{fig_Shafranov_shift}
\end{figure}
\noindent As to the boundary conditions, these have been obtained in the following way:

\noindent {\bf a)} First of all, we have to satisfy the Onsager condition. Hence, the solution should vanish at the origin of the axes, i.e., $\phi(\mathbf{0}) =0$;

\noindent {\bf b)} Experimental evidences show that, in pure collisional regime, the pure effects (such as Fourier's law, Fick's law etc.) are very robust laws. So, we have to impose $\phi(X^1=0,X^2)=\phi(X^1,X^2=0)=0$;

\noindent {\bf c)} There are no privileged directions when the thermodynamic forces tend to infinity (or for very large values of the thermodynamic forces). In other words, $\phi(r=R_0,\theta)=const.\equiv c_0\neq 0$ on the arc of a circle of radius $R_0$ (with $R_0$ very large). Here, $(r,\theta)$ denotes the polar coordinates: $r=({X^1}^2+{X^2}^2)^{1/2}$ and $\theta=\arctan(X^{2}/X^{1})$.

\noindent The boundary conditions, in case of FTU-plasmas in fully collisional regime are depicted in Fig.~(\ref{fig_BC1}). 
\begin{figure}[b]
\sidecaption
\includegraphics[scale=.60]{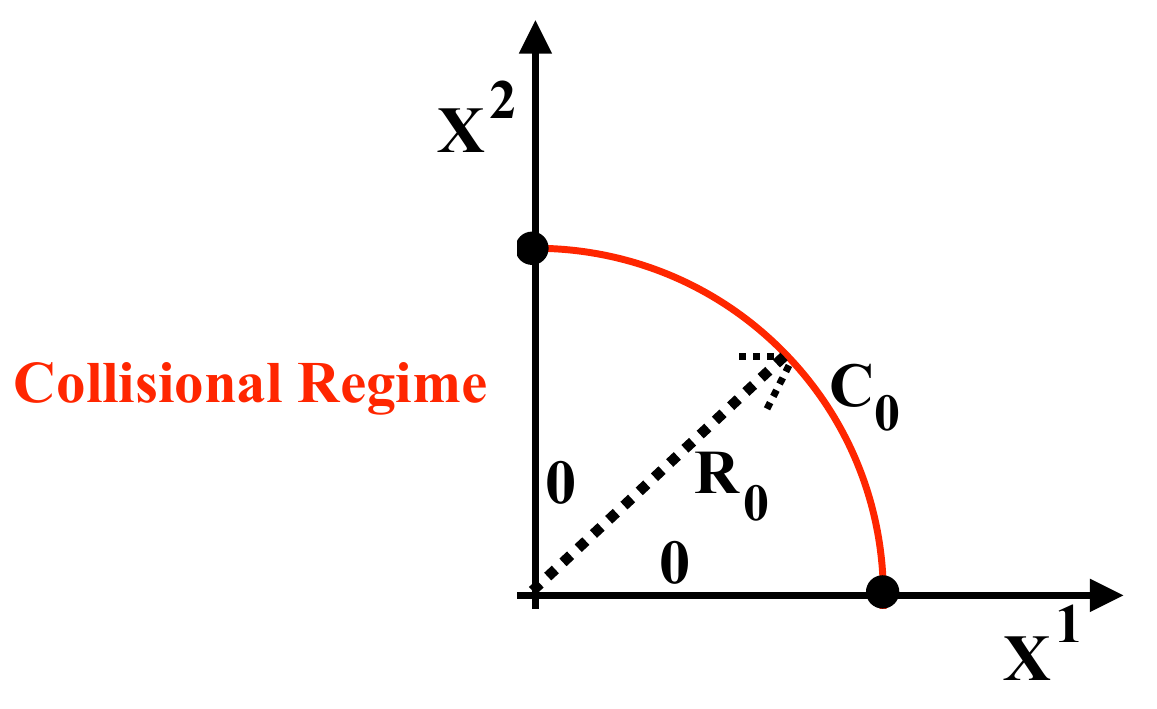}
\caption{{\bf Boundary conditions for $\phi$ for collisional FTU-plasmas}. The validity of Onsager's relations is ensured by imposing that $\phi$ vanishes along the axes. In addition, we have to impose that there are no privileged directions for very large values of the thermodynamic forces. This condition is satisfied by imposing, in the first quadrant, $\phi(r=R_0,\theta)=c_0\neq0$.} 
\label{fig_BC1}
\end{figure}

\noindent Now, we are able to solve the PDE~(\ref{conf10}) in the first quadrant. After having obtained the solution in the first quadrant, successively we shall be able to reconstruct the entire solution which is valid for the whole circle by using the Schwartz principle \cite{CourantHilbert1}. Parameters $R_0$ and $c_0$ have been determined as follows. 

\noindent The scaling parameter $R_0$ is determined such that 

\noindent {\bf i)} a solution of the TFT equation exists everywhere in the physical system, hence it cannot be too small;

\noindent {\bf ii)} the solution area is maximised in the space of the thermodynamic forces, i.e. $R_0$ defines the minimal circle enclosing the solution area - see Fig.~(\ref{R0}).
\begin{figure}[b]
\sidecaption
\includegraphics[scale=.30]{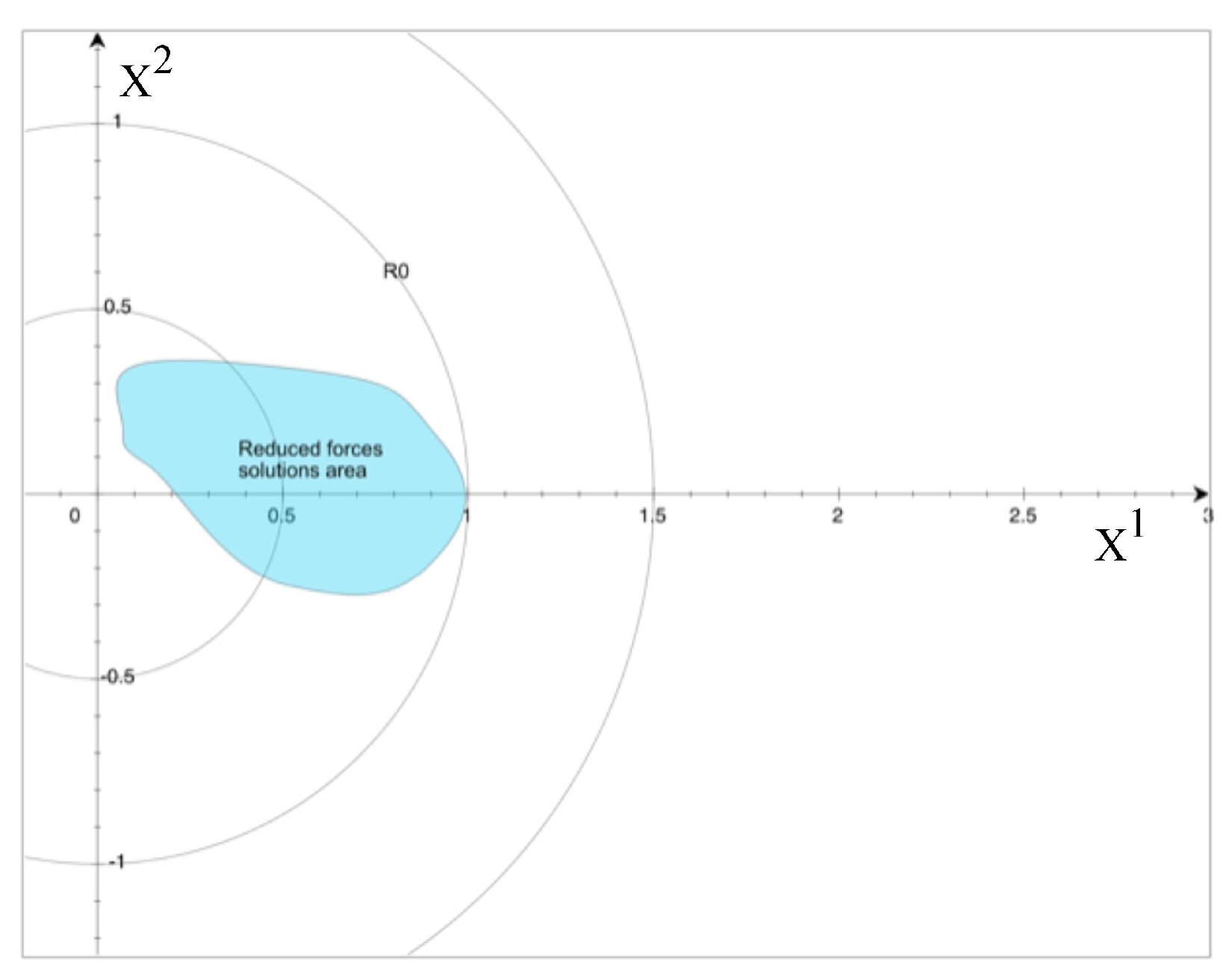}
\caption{{\bf Determination of the value $R_0$}. $R_0$ defines the minimal circle enclosing the solution area. Here, the solution area is maximised in the thermodynamic forces space, spanned by $X^1$ and $X^2$.}
\label{R0}
\end{figure}

\noindent The Dirichlet boundary condition $c_0$ is determined such that the thermodynamic forces $X^1$ and $X^2$, solutions of the system, maximise the electron heat loss. It is numerically found to be approximately equal to $c_0\simeq -4.5$.

\noindent Fig.~(\ref{fig_TFT_Exp_Theory_Comparison}) shows a comparison between experimental data for fully collisional FTU-plasmas and the theoretical predictions of Eq.~(\ref{conf10}), subjected to the boundary conditions illustrated in Fig.~(\ref{fig_BC1}). In the vertical axis we have the (surface magnetic-averaged) radial electron heat flux, and in the horizontal axis the minor radius of the tokamak. The lowest dashed profile corresponds to the Onsager theory (i.e., the neoclassical theory) and the bold line to the \textit{Thermodynamical Field Theory} (TFT)} satisfying the TCP, respectively. The highest profile is the experimental data provided by the ENEA C.R.-EUROfusion. As we can see, the TCP principle is well satisfied in the core of the plasma where plasma is in the fully collisional transport regime. Towards the edge of the tokamak, transport is dominated by turbulence.

\begin{figure}[b]
\sidecaption
\includegraphics[scale=.65]{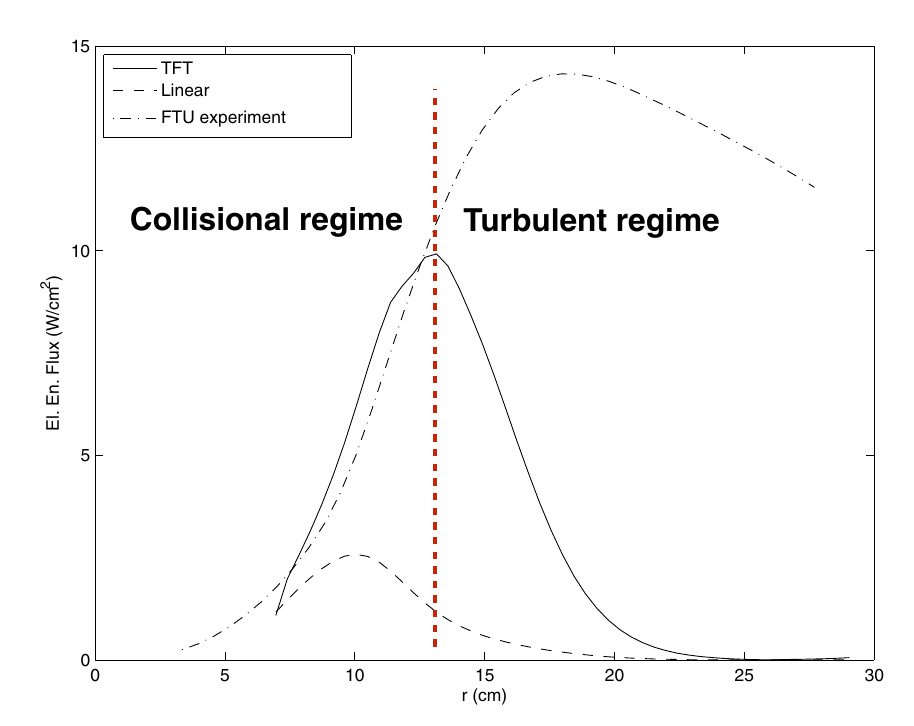}
\caption{{\bf Electron heat loss in FTU-plasmas vs the minor radius of the Tokamak}. The highest dashed line is the experimental profile. These data have been provided by Marinucci from the ENEA C.R.-EUROfusion in Frascati \cite{marinucci}. The bold line is the theoretical profile obtained by the nonlinear theory satisfying the TCP (TFT) and the lowest dashed profile corresponds to the theoretical prediction obtained by Onsager's theory (i.e., by the noclassical theory).}
\label{fig_TFT_Exp_Theory_Comparison}
\end{figure}

\noindent Fig.~(\ref{fig_TFT_Turbulent}) shows a comparison between experimental data and the theoretical predictions for FTU-plasmas in collisional as well as in turbulent regimes. The boundary conditions to be satisfied by Eq.~(\ref{conf10}), which are valid in the collision zone as well as in the turbulent one are illustrated in Fig.~(\ref{fig_BC2}). Mathematical details related to the solutions of the equations for FTU-plasmas in turbulent regime can be found in a manuscript recently submitted to publication in the review {\it Chaos Solitons and Fractals} (2022). As we can see, the agreement between the experimental data and the theoretical predictions is fair both qualitatively and quantitatively. 

\begin{figure}[b]
\sidecaption
\includegraphics[scale=.30]{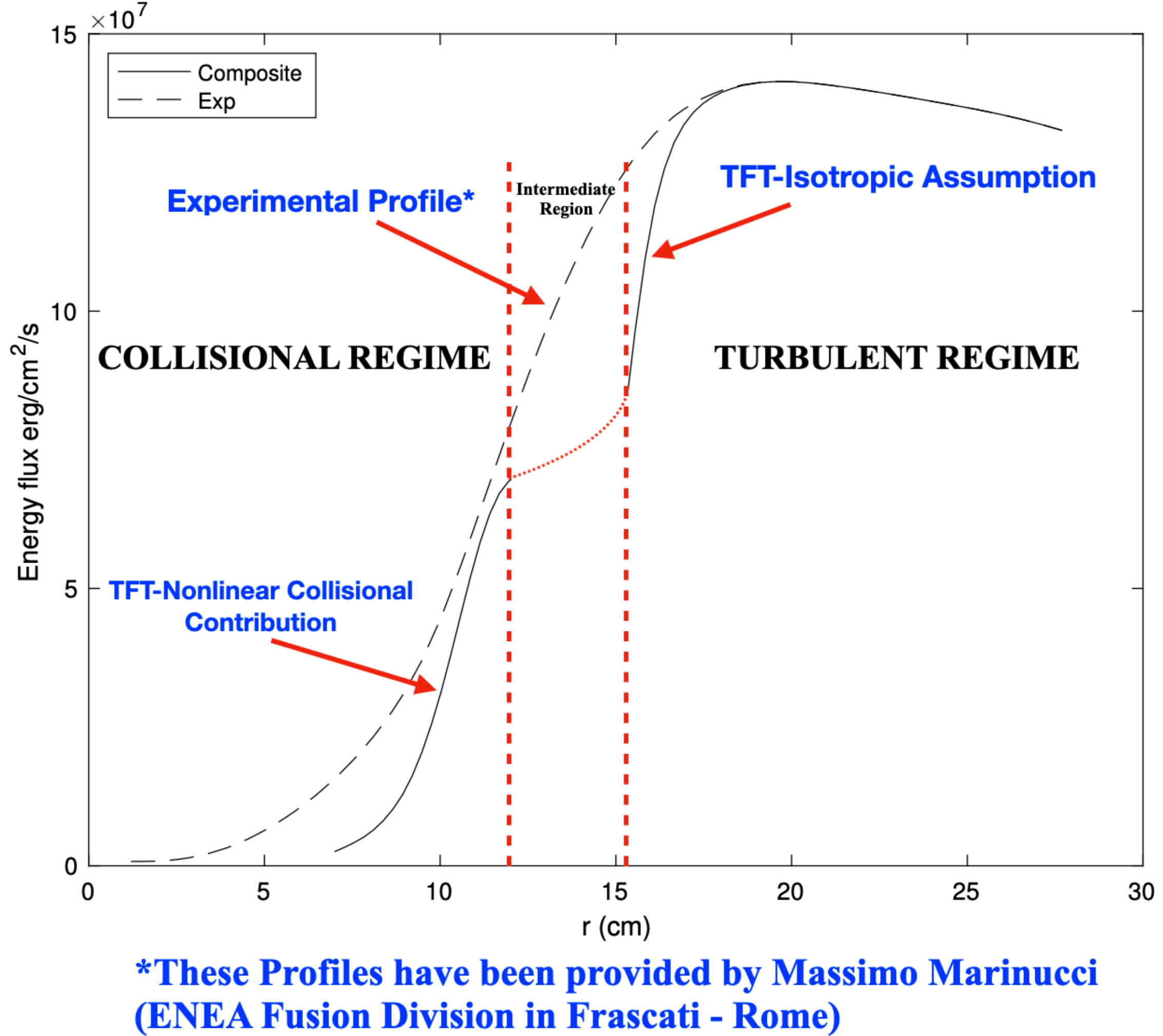}
\caption{{\bf Electron heat loss in FTU-plasmas in turbulent regime}. The highest dashed line is the experimental profile. These data have been provided by Marinucci from the ENEA C.R.-EUROfusion in Frascati \cite{marinucci}. The bold line refers to the theoretical predictions obtained by solving the TFT-equations for FTU-plasmas in collisional regime (plasma is near the center of the Tokamak) and in turbulent regime (plasma is close to the boundary of the Tokamak).}
\label{fig_TFT_Turbulent}
\end{figure}


\section{Conclusions}\label{conclusions}
A non-Riemannian geometry has been constructed out of the components of the affine connection which has been determined by imposing the validity of the \textit{General Evolution Criterion} for non-equilibrium systems relaxing toward a steady state. Relaxation expresses an intrinsic physical property of a thermodynamic system. The affine connection, on the other hand, is an intrinsic property of geometry allowing to perform derivatives and to determine the equation for the shortest path. It is spontaneous to argue that a correct thermodynamical-geometrical theory should correlate these two properties: relaxation of a system with the affine connection. It is important to recall that the General Evolution Criterion is valid for systems, even far, from equilibrium and even in turbulent regime. More specifically, this theorem has been derived only from the balance equations for mass, energy and momentum \textit{without assuming the validity of the Onsager reciprocity relations and without neglecting any terms, including the terms leadings to turbulence in hydrodynamic systems}. Successively, Glansdorff and Prigogine showed the validity of this theorem also for plasmas governed by the plasma-dynamic PDEs. Incidentally, if we assume that the transport coefficients are a small perturbation of the Onsager matrix and in the limit $\sigma >>1$, all terms leading to turbulence disappear and the General Evolution Criterion is trivially satisfied. In this limit case, we obtain the closure relations reported in ref.~\cite{sonnino1}. 

\noindent \textit{Action~(\ref{TAP6}) (or action~(\ref{TAP2})), takes into account all the terms of the balance equations, including those leading to turbulence, and its validity range coincides with that of the General Evolution Criterion}. The action principle leads, for $n>2$, to the PDEs~(\ref{nte7}) and, for $n=2$, to Eq.~(\ref{conf10}), respectively. To get these equations we did not neglect any terms and we did not require that the transport coefficients are close to the Onsager matrix. To investigate turbulence, we did not assume that $\sigma >>1$. Successively, we have applied Eq.~(\ref{conf10}) to FTU-plasmas in collisional regime. This regime requires that the \textit {pure Onsager laws} (i.e. the Fourier law, the Fick law etc.) are \textit{very robust laws} and for this we are bound to impose as boundary conditions that along the thermodynamic axes the transport coefficients must coincide with the Onsager ones. It is worth noticing that variable $\phi$ is not a perturbation. It is our opinion that it is a great success that in the collision regime the theoretical predictions, resulting from a PDEs so different from the standard equations that we are used to seeing in literature, are in very good agreement with the experimental data. Since the PDE~(\ref{conf10}) has been derived without neglecting any term present in the dynamic equations (i.e. the energy, mass and momentum balance equations), it is quite natural to propose Eq.~(\ref{conf10}) as a good candidate also for describing transport in two-dimensional turbulent systems. We have also investigated the electron heat loss for FTU-plasmas even in the turbulent zone by specifying the appropriate boundary conditions. In the turbulent zone, the system is (very) far from thermodynamic equilibrium. Thus, in this zone we released the very strict condition that along the (thermodynamic) axes the solution must coincide with the Onsager relations. Indeed, in turbulent regime the Onsager regression hypothesis for microscopic fluctuations of small non-equilibrium disturbances is violated \cite{onsager1}. The boundary conditions, in case of Tokamak-plasmas in the collisional regime (first circle) with the ones in the turbulent regime (i.e., in the annulus) are depicted in Fig.~(\ref{fig_BC2}). Concretely, we determined the conditions where the constant solution of Eq.~(\ref{conf10}) loses its stability towards a new one. This task has been accomplished by applying, for example, the mathematical methods reported in ref.~\cite{awrejcewicz}. The ultimate aim of our work is to apply our approach to the \textit{Divertor Tokamak Test facility} (DTT) to be built in Italy and to ITER.

\noindent We conclude with some comments about the validity of Eq.~(\ref{I1}). It is known that the most general flux-force transport relations takes the form
\begin{equation}\label{r1}
J_\mu({\bf r},t)=\int_{\Omega}d{\bf r}'\int_0^t dt'{\mathcal G}_{\mu\nu}[X({\bf r}',t')]X^{'\!\nu}({\bf r}-{\bf r}',t-t')
\end{equation}
\noindent with $\Omega$ denoting the volume occupied by the system. The space-time dependent coefficients ${\mathcal G}_{\mu\nu}$ are called \textit{nonlocal transport coefficients}: they should not be confused with coefficients $\varpi_{\mu\nu}$ (they do not have the same dimension). The nonlocal and non-Markovian Eq.~(\ref{r1}) expresses the fact that the flux at a given point $({\bf r},{\rm t})$ could be influenced by the values of the forces in its spatial environment and by its history. Whenever the spatial and temporal ranges of influence are sufficiently small, the delocalisation and the retardation of the forces can be neglected under the integral,
\begin{align}\label{r2}
&{\mathcal G}_{\mu\nu}[X({\bf r}',t')]X^{'\!\nu}({\bf r}-{\bf r}',t-t')\\
&=2\varpi_{\mu\nu}[X({\bf r},t)]X^\nu({\bf r},t)\delta({\bf r}-{\bf r}')\delta(t-t')\nonumber
\end{align}
\noindent with $\delta$ denoting Dirac's delta function. In this case, the transport equations reduces to
\begin{equation}\label{r3}
J_\mu({\bf r},t)=\varpi_{\mu\nu}[X({\bf r},t)]X^\nu({\bf r},t)
\end{equation}
\noindent In the vast majority of cases studied at present in transport theory, it is assumed that the transport equations are of the form of Eq.~(\ref{r3}). However, equations of the form~(\ref{r1}) may be met when we deal with anomalous transport processes such as, for example, transport in turbulent tokamak plasmas - see, for example, ref.~\cite{balescu1}. Hence, Eqs~(\ref{r2}) establish, in some sort, the limit of validity of Eq.~(\ref{I1}) and, in this case, the fluxes should be evaluated by using Eq.~(\ref{r1}). Nonetheless, we would like to stress the following. Hydrodynamic turbulence is normally studied through the Navier-Stokes equations, supported by the conservation equations for the mass and energy (the so-called \textit{mass-energy balance equations}). The set of hydrodynamic equations are closed through relations of the form~(\ref{r3}) where, for Newtonian fluids, $\tau_{\mu\nu}$ depends only on the thermodynamical quantities, and not on their gradients. The experimental data are in excellent agreement with the numerical simulations - see, for example, ref.~\cite{kollmann}. For non-Newtonian fluids, turbulence is still analysed by closing the balance equations with equations of the form~(\ref{r3}) where the viscosity coefficients depends not only on the thermodynamic quantities but also on their gradients - see, for example, ref.~\cite{lumley}. Also in this case, the experimental data are in excellent agreement with the numerical simulations. Even transport phenomena in Tokamak-plasmas in the weak-collisional regime are analysed by closing the balance equations with equations of the type~(\ref{r3}) - see, for example, ref.~\cite{balescu2}. This is for saying that Eqs~(\ref{r3}) are \textit{very robust} equations and their validity goes well beyond the collisional, or the weak-collisional, regime. This case is very similar to what happens for the Onsager reciprocity relations: even if, according to the non-equilibrium statistical physics and the kinetic theory, these relations should have been valid only in vicinity of the thermodynamic equilibrium in reality their validity goes well beyond the thermodynamic equilibrium, up to be valid even in turbulent hydrodynamic regimes. 

\noindent In conclusion, before further complicating the mathematical formalism, it is the author's opinion that it is still worth analysing the turbulence in Tokamak plasmas by closing the balance equations with local equations of the type~(\ref{r3}) and comparing \textit{a posteriori} the theoretical predictions with the experimental data.

\noindent By passing, there is another important point which is worthwhile mentioning. In this manuscript, the thermodynamic quantities (number density, temperature, pressure, etc.) are evaluated at the local equilibrium state. This is not inconsistent with the fact that the arbitrary state of a thermodynamic system is close to, but not in a state of local equilibrium. Indeed, as known, it is always possible to construct a representation in such a way that the thermodynamic quantities evaluated with a distribution function close to a Maxwellian do coincide exactly with those evaluated at the local equilibrium state - see, for example, the textbook~\cite{balescu3}. 

\begin{figure}[b]
\sidecaption
\includegraphics[width=5.5cm]{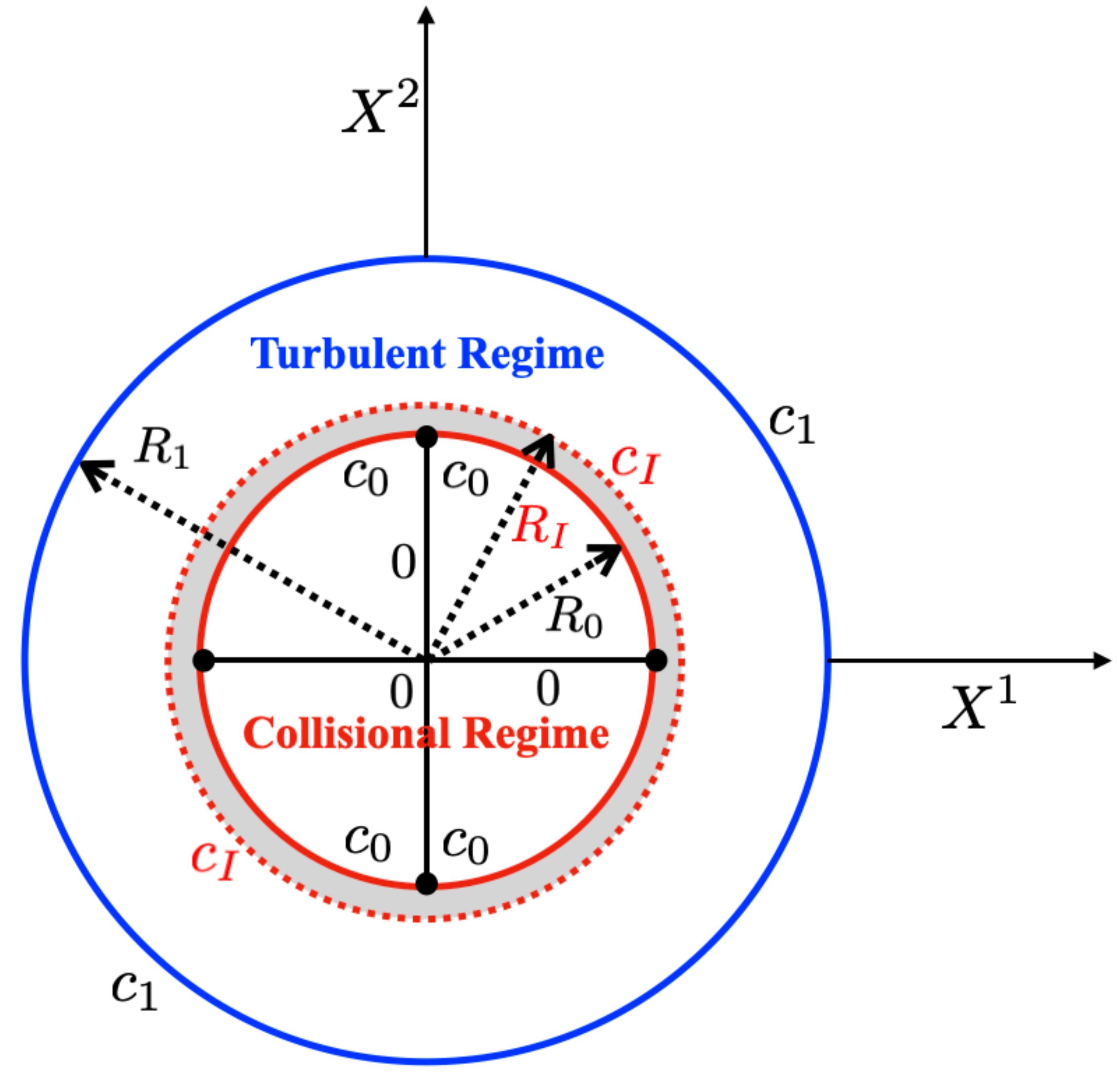}
\caption{{\bf Boundary conditions for $\phi$ for FTU-plasmas in collisional and in turbulent regimes}. In the turbulent zone, plasmas is far from equilibrium. So, in the annulus the Onsager relations are no longer valid. Of course, we have always to impose that there are no privileged directions for very large values (say for $r=R_1$) of the thermodynamic forces. This condition is satisfied by imposing, in the first quadrant, $\phi (r = R_1,\theta)=c_1\neq 0 $. By combining these boundary conditions with those specified for FTU-plasmas in the collision regime, we are finally able to find the solution of Eq.~(\ref{conf10}), valid in the collision zone as well as in the turbulent one.}
\label{fig_BC2}
\end{figure}

\noindent {\bf Data Availability Statement}

\noindent The data that support the findings of this study are available from the corresponding author
upon reasonable request.

\noindent  {\bf Acknowledgements}

\noindent I pay tribute to my friend, Prof. Slava Belyi. I remember fondly, and emotion, the fruitful discussions, at times animated, on issues concerning Nonequilibrium Thermodynamics at home of Prof. Ilya Prigogine. Prof. I. Prigogine often invited us to dinner at his home and I remember that the discussions soon ended in converging on topics concerning the dynamics of systems out of equilibrium, the role of fluctuations and the formulation of the sub-dynamics. 

\noindent I would like also to pay tribute to my colleague and friend Prof. Enrique Tirapegui, co-author of several manuscripts of this series of works.

\noindent I am indebted to Dr M. Marinucci from the ENEA - Frascati (Rome-Italy) for having provided the experimental data for the FTU-plasmas.

\section*{Appendix}
\addcontentsline{toc}{section}{Appendix}

\noindent$\bullet$ {\bf Specification of the Boundary Conditions}
\vskip 0.2truecm
\noindent The purpose of this Section is to specify the boundary conditions for the first equation of system~(\ref{new12}), i.e., for the Laplace PDE
\begin{equation}\label{laplace1}
{\rm I}^{\lambda\kappa}\frac{\partial^2f(x)}{\partial X^{\lambda}\partial X^{\kappa}}=0
\end{equation}
\noindent with ${\rm I}^{\lambda\kappa}$ denoting the components of the Identity matrix. This task will be accomplished by tacking into account the Onsager theory and experimental evidences. More specifically, we should require the validity of the following conditions:
\begin{description}
\item[a)] The solution must coincide with the Onsager matrix when the system approaches equilibrium. We refer this condition to as {\it Onsager's condition};
\item[b)] Experimental evidences show that the pure effects (such as Fourier's law, Fick's law etc.) are very robust laws. Hence, for the unidimensional case (i.e., for $n=1$), we impose $g_{11}=L_{11}$ (or $h_{11}=0$);
\item[c)]There are no privileged directions when the thermodynamic forces tend to infinity (or for very large values of the thermodynamic forces);
\item[d)] For isotropic substances, the solution must be invariant under permutation among the (dimensionless) thermodynamic forces. Hence, for isotropic materials, the solution should be invariant with respect to the permutation of the axes $X^i$;
\item[e)]  The solution holding throughout the space may be obtained by using the Schwartz principle \cite{CourantHilbert1};
\item[f)] The boundary conditions for the $n$-dimensional case may be derived from the knowledge of the solution of Eq.~(\ref{laplace1}) for the $(n-1)$-dimensional case.
\end{description}
\noindent The analysis of the two-dimensional case will made clear this approach. Once solved this case, we shall be able to specify the boundary conditions for a $3$-dimensional thermodynamic space and so on.
\begin{description}
\item[i)] First of all we have to satisfy the Onsager condition. Hence, the solution should vanish at the origin of the axes: 
\begin{equation}\label{a}
f(\mathbf{0}) =0
\end{equation}
\noindent As a consequence, in two dimensions, we should have $f(r=0,\theta)=0$, with $r=({X^1}^2+{X^2}^2)^{1/2}$ and $\theta=\arctan(X^{2}/X^{1})$, respectively. 
\item[ii)] From condition {\bf b)} we have that the perturbation of the transport coefficients is zero on the axes $X^1=0$ and $X^2=0$, and condition {\bf c)} requires that the solution should be a constant different from zero, say with value $k\neq 0$, on the arc of a circle of radius $R_0$ (with $R_0$ very large). Hence, we should have:
\begin{equation}\label{b}
f(\mathbf{X}){\mid}_{(\theta=0, r)}=f(\mathbf{X}){\mid}_{(\theta=\pi/2, r)}=0\ \ {\rm and}\ \  f(\mathbf{X}){\mid}_{(0<\theta <\pi/2, r=R_{0})}=k\neq 0
\end{equation}
\item[iii)] The solution should be invariant with respect to the permutation of the axes $X^1$ and $X^2$ ;
\item[iv)] After having obtained the solution in the first quadrant, successively we shall be able to reconstruct the entire solution which is valid for the whole circle by using the Schwartz principle \cite{CourantHilbert1}. 
\end{description}
\noindent Note that, according to the previous boundary condition {\bf b)}, the first derivatives of the solution have discontinuity points. However, the $\mathit{C^{2}}$  smoothness inside the domain is automatically assured by Weyl's lemma \cite{weyl}. Hence, the solutions are of of class $\mathit{C^{2}}$ inside the circle, except at the boundary where they are at least of class $\mathit{C^{0}}$. 

\noindent By tacking into account conditions {\bf i)}-{\bf iv)}, it is easy to convince ourselves that, for $n=2$, the correct boundary condition read \cite {sonnino4} and \cite {sonnino3}:
\begin{equation}\label{circle}
f(R_{0},\theta) = \left\{ \begin{array}{ll}
k & \mbox{if $ 0 < \theta<\frac{\pi}{2}$}\\
-k & \mbox{if $\frac{\pi}{2} < \theta < \pi$}\\
k & \mbox{if $ - \pi < \theta< - \frac{\pi}{2}$}\\
-k & \mbox{if $- \frac{\pi}{2} < \theta< 0$} 
\end{array}
\right.
\end{equation}
\vskip 0.2truecm
\noindent$\bullet$ {\bf Solution of Eq.~(\ref{laplace1}) for the Two-Dimensional Case}
\vskip 0.2truecm
The solution of this problem can be found in \cite {sonnino4} and \cite {sonnino3}. Here, we shall solve Eq.~(\ref{laplace1}) subject to the boundary conditions depicted in Fig.~(\ref{2D}).
\begin{figure}[b]
\sidecaption
\includegraphics[scale=.60]{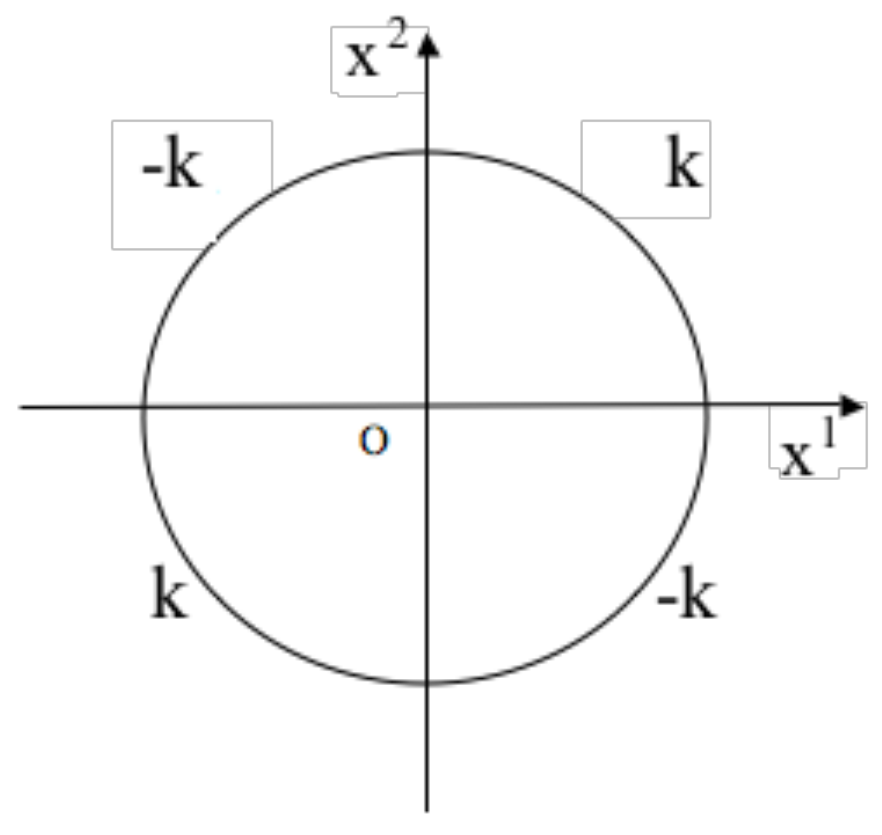}
\caption{ The 2-dimensional boundary conditions.}
\label{2D}
\end{figure}

\noindent As known, the solution of Eq.~(\ref{laplace1}) with boundary conditions (\ref{circle}) can be written in general as (see, for example \cite{zachmanoglou} or \cite{morse}):
\begin{equation}\label{circle1}
f(r,\theta) = \frac{a_{0}}{2}+\sum_{n=1}^{\infty}\ (a_{n}\cos n\theta  +b_{n}\sin n\theta )\ \biggl (
\frac{r}{R_{0}}\biggr )^{n}
\end{equation}
\noindent with
\begin{eqnarray}\label{circle2}
a_{0} & = & 2\int_{-\pi}^{\pi}f(R_{0},\theta)\ d\theta\nonumber \\
a_{n} & = & \frac{1}{\pi}\int_{-\pi}^{\pi}f(R_{0},\theta)\cos (n\theta )\ d\theta\qquad
\qquad\mathrm{for}\qquad\mathrm{n}\geq 1 \\
b_{n} & = & \frac{1}{\pi}\int_{-\pi}^{\pi}f(R_{0},\theta)\sin (n\theta )\ d\theta \qquad
\qquad\mathrm{for}\qquad\mathrm{n}\geq 1 \nonumber
\end{eqnarray}
\noindent Integrals~(\ref{circle2}) can be computed. We get
\begin{eqnarray}\label{circle3}
a_{n} & = & 0\qquad\qquad\qquad\qquad\qquad\qquad\quad\ \ \mathrm{(n=0,1,2}\cdots )\\
b_{n} & = & -k\frac{8\cos (\frac{n\pi}{2})\sin (\frac{n\pi}{4})^2}{n\pi}\qquad\qquad\ \ \mathrm{(n=1,2}\cdots )\nonumber
\end{eqnarray}
\noindent Therefore solution $f(r,\theta)$ can be written as
\begin{equation}\label{circle4}
f(r,\theta) = -\frac{8 k}{\pi}\sum_{n=1}^{\infty}\frac{\cos (\frac{n\pi}{2})\sin (\frac{n\pi}{4})^2}{n}\ \sin (n\theta )\biggl (\frac{r}{R_{0}}\biggr )^{n}
\end{equation}
\noindent Solution (\ref{circle4}) can be brought into the form
\begin{equation}\label{circle5}
f(r,\theta) = \frac{4 k}{\pi}\sum_{n=1}^{\infty}\frac{\sin (2(n-1)\theta)}{2n-1}\biggl (\frac{r}{R_{0}}\biggr )^{n}
\end{equation}
\noindent The sum in Eq.~(\ref{circle5}) can be evaluated \cite{gradshteyn}, and we find the compact expression 
\begin{equation}\label{circle6}
f(r,\theta) = \frac {2 k}{\pi}\arctan \biggl [\frac{2{\rho}^2\sin (2\theta )}{1-{\rho}^4}\biggr ]\quad\qquad\mathrm{where}\quad{\rho\equiv\frac{r}{R_{0}}}\leq 1
\end{equation}
\noindent or, in coordinate $x^1$ and $x^2$:
\begin{equation}\label{circle7}
f(X^1, X^2) = \frac {2 k}{\pi}\arctan \biggl [\frac{4R_{0}^2 {X^1}{X^2}}{R_{0}^4-({X^1}^{2}+{X^2}^{2})^2}\biggr ]
\end{equation}
\noindent Solution (\ref{circle7}) is valid only in the quadrants $X^1X^2 > 0$. The general solution, valid in all quadrants, is obtained by using Schwartz's principle:
\begin{equation}\label{circle8}
f(X^1, X^2) = \frac {2 k}{\pi}\arctan \biggl [\frac{4R_{0}^2 \mid X^1 X^2\mid}{R_{0}^4-({X^1}^{2}+{X^2}^2)^2}\biggr ]
\end{equation}
\noindent In our original problem, we found that constant $k=1$ \cite{sonnino9}. Solution~(\ref{circle8}) is illustrated in Fig.~(\ref{Solution2D}).
\begin{figure}[b]
\sidecaption
\includegraphics[width=6.0cm,height=6.0cm]{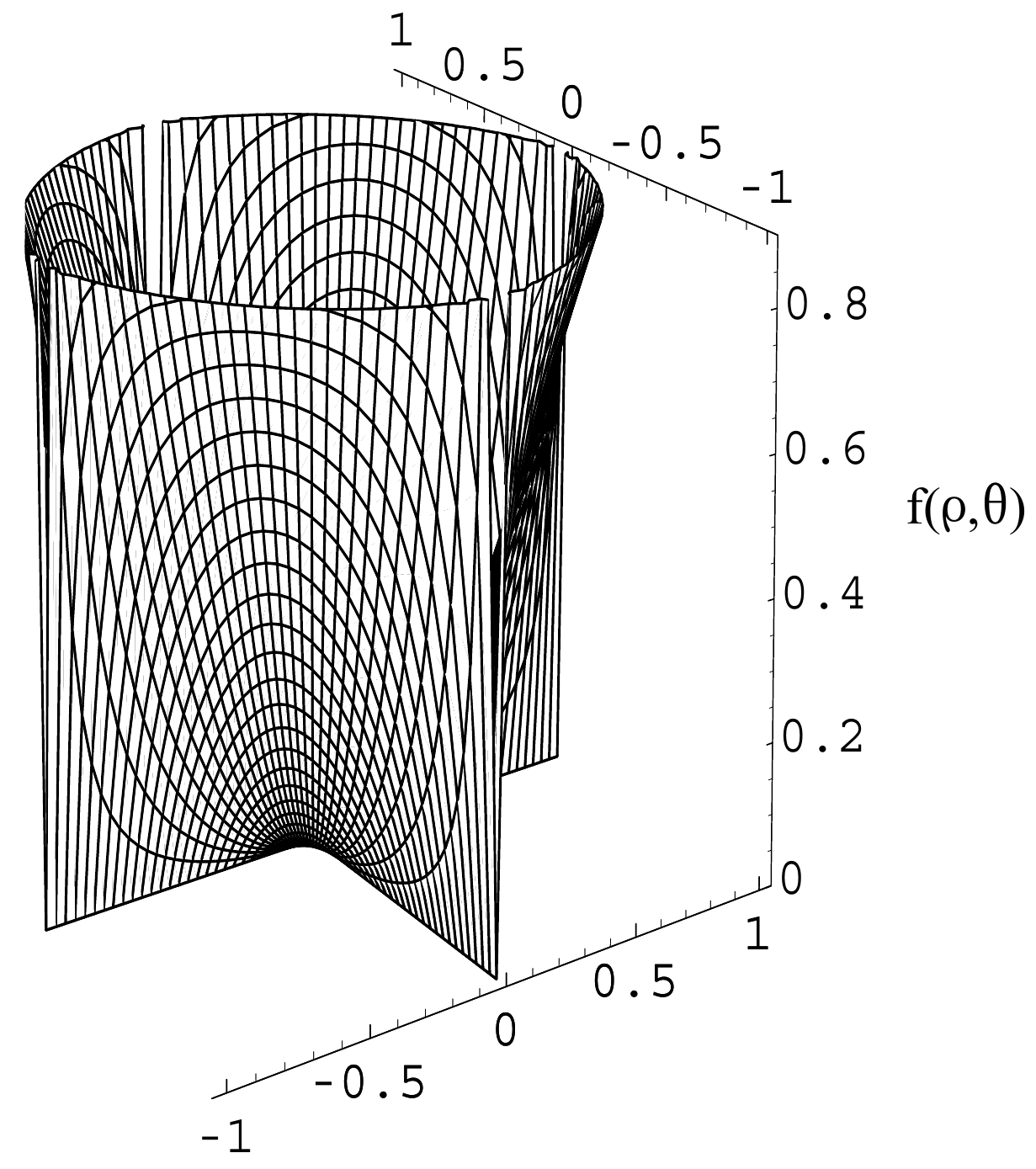}
\caption{ \label{Solution2D} Solution~(\ref{circle8}) (with $k=1$) in polar coordinates. }
\end{figure}
\vskip 0.2truecm
\noindent$\bullet$ {\bf Boundary Conditions for the 3-Dimensional Space}
\vskip 0.2truecm
The boundary conditions for a three dimensions space are derived directly from the solution of Eq.~(\ref{laplace1}) in two dimensions. Indeed, let us consider the first octant of the space. In the planes $X^3=0$, $X^2=0$ and $X^1=0$ we should re-obtain the expressions for the transport coefficients which we have found by solving the two-dimensional case. So, on the planes $X^1=0$, $X^2=0$ and $X^3=0$, solution $f(X^1, X^2, X^3)$ should satisfies the boundary conditions 
\begin{eqnarray}\label{3bc1}
f(x^1=0, X^2, X^3)=f_1(X^2, X^3) &=& \frac {2 k}{\pi}\arctan \biggl [\frac{4R_{0}^2 {X^2}{X^3}}{R_{0}^4-({X^2}^{2}+{X^3}^{2})^2}\biggr ]\\
f(X^1, X^2=0, X^3)=f_2(X^1, X^3) &=& \frac {2 k}{\pi}\arctan \biggl [\frac{4R_{0}^2 {X^1}{X^3}}{R_{0}^4-({X^1}^{2}+{X^3}^{2})^2}\biggr ]\nonumber\\
f(X^1, X^2, X^3=0)=f_3(X^1, X^2) &=& \frac {2 k}{\pi}\arctan \biggl [\frac{4R_{0}^2 {X^1}{X^2}}{R_{0}^4-({X^1}^{2}+{X^2}^{2})^2}\biggr ],\nonumber
\end{eqnarray}
\noindent respectively. In addition, the above condition {\bf d)} is satisfied by imposing that the solution $f(r,\theta,\phi)$ (with $r$, $\theta$ and $\phi$ denoting the radial coordinate, the azimuth, and the zenith angle, respectively) is $constant$ on the spherical cap of radius $R_0$, centred at the origin of the axes and located in the first octant:
\begin{equation}\label{3bc2}
f(r=R_0,\theta,\phi)=k\neq 0
\end{equation}
\noindent In this way, the above-conditions {\bf a)}-{\bf d)} have been satisfied and we have obtained a well-posed Dirichlet's problem in the first octant. Now, we are in a position to solve the Laplace PDE in the first octant, subject to the above-derived Dirichlet's boundary conditions, by using standard methods of mathematical physics\footnote{See, for example, the reference books \cite{morse}, \cite{whittaker}, \cite{CourantHilbert2}.}. As for the two-dimensional case, the solution holding throughout the space may be obtained by using the Schwartz principle. Of course, the above method is not limited to the three-dimensional case, and it can naturally be extended for getting the boundary conditions for a $n$-dimensional space, once obtained the solution of Eq.~(\ref{laplace1}) for a $(n-1)$-dimensional space.
\vskip 0.2truecm
\noindent$\bullet$ {\bf Solutions of Poisson's PDE}
\vskip0.2truecm

\noindent In this Section we solve the Poisson PDE subjected to the appropriate boundary conditions,
\begin{equation}\label{app21}
{\rm I}^{\lambda\kappa}\frac{\partial h^{(1)}(X)}{\partial X^\lambda\partial X^\kappa}=W^{(S)}(X)
\end{equation}
\noindent with source $W^{(S)}(X)$ given by the right-hand side of the second equation in system~(\ref{new12}). According to the general procedure, we have, firstly, to find a particular solution of the Poisson PDE which should be solved with all homogeneous boundary conditions. The individual conditions must retain their type (Dirichlet, Neumann or Robin type) in this sub-problem. Successively, we have to add any solution of the homogeneous Laplace equation with the non-homogeneous boundary conditions. Also in this case, the individual conditions must retain their type (Dirichlet, Neumann or Robin type) in the sub-problem. The complete solution of the Poisson equation is the sum of the solution of the two sub-problems: the solution of the {\it Poisson sub-problem} plus the solution of the {\it Laplace sub-problem}. Since the boundary conditions have already been satisfied when we solved the PDE at the dominant order (i.e., the equations for $h_{\mu\nu}^{(0)}$), the only task that we have to accomplish is to find the particular solution of the inhomogeneous Poisson PDE subject to homogeneous boundary conditions. The solution to the homogeneous equation allows us to obtain a system of basis functions that satisfy the given boundary conditions. 

\noindent For a two-dimensional space (i.e., in case of two independent thermodynamic forces), Eq.~(\ref{circle5}) suggests the following expression for the solution of the Poisson equation
\begin{equation}\label{app22}
h^{(1)}(r,\theta) = \sum_{n=1}^{\infty} {a_n}(r)\sin \left(2(n-1)\theta\right)
\end{equation}
\noindent Eq.~(\ref{app22}) satisfies the homogeneous boundary conditions at $\theta=0$ and $\theta=\pi/2$. Now, we substitute Eq.~(\ref{app22}) into the Poisson equation~(\ref{app21}), written in polar coordinates, i.e.:
\begin{equation}\label{app23}
\frac{1}{r}\frac{\partial}{\partial r}\Bigl(r\frac{\partial h^{(1)}}{\partial r}\Bigr)+\frac{1}{r^2}\frac{\partial^2 h^{(1)}}{\partial\theta^2}= W^{(S)}(r,\theta)
\end{equation}
\noindent and we solve the equation by using the orthogonality relations for the sine functions. We have also to take into account that, at equilibrium (i.e. at $\rho=0$), we have to re-obtain the Onsager matrix and we have also to satisfy the homogeneous condition at $\rho=1$. Finally, after simple calculations, we get the following ordinary differential equation for $a_n(r)$, subject to the following conditions:
\begin{eqnarray}\label{app24}
&&\rho^2 {a''_n}(\rho)+\rho {a'_n}(\rho)-4(n-1)^2{a_n}(\rho)=R_0^2\rho^2 {\widehat W}^{(S)}_n\!(\rho)\\ 
&&{a_n}(0)=0\ \ ;\ \ {a_n}(1)=0\qquad{\rm where}\nonumber\\
&&{\widehat W}^{(S)}_n\!(\rho)\equiv\frac{1}{\pi}\int_{-\pi}^{\pi}W^{(S)}(\rho,\theta)\sin\left(2(n-1)\theta\right)d\theta\nonumber
\end{eqnarray}
\noindent with $\rho\equiv r/R_0\leq 1$ and the suffix "prime" denoting the derivative with respect to $\rho$, respectively. Eq.~(\ref{app24}) corresponds to a {\it non-homogeneous Euler equation of the $2^{nd}$-order.} By using standard methods of integration (see, for example \cite{walter}), by imposing the Onsager and the homogeneous conditions, after simple calculations we finally get 
\begin{eqnarray}\label{app25}
a_n(\rho)&=&-\frac{R_0^2}{4(n-1)}\Bigl(\rho^{-2(n-1)}\int_0^\rho t^{2n-1}{\widehat W}^{(S)}_n\!(t)dt\\
&-&\rho^{2(n-1)}\int_0^\rho t^{-2n+3}{\widehat W}^{(S)}_n\!(t)dt\Bigr)+\alpha_{n}\rho^{2(n-1)}\quad {\rm with}\nonumber\\
\alpha_{n}&=&-\frac{R_0^2}{4(n-1)}\int_0^1\left(t^{-2n+3}-t^{2n-1}\right){\widehat W}^{(S)}_n\!(t)dt\nonumber
\end{eqnarray}
\noindent Solution~(\ref{app25}) satisfies the Onsager condition since\footnote{The indeterminate form can be solved by using Hospital's rule and $\lim_{\rho\rightarrow 0} {\widehat W}_n^{(S)}(\rho)=0$.}
\begin{equation}\label{app26}
\lim_{\rho\rightarrow 0}\rho^{-2(n-1)}\int_0^\rho t^{2n-1}{\widehat W}^{(S)}_n\!(t)dt=0
\end{equation}
\noindent Hence,
\begin{eqnarray}\label{app27}
h^{(1)}(\rho,\theta)&=&R_0^2\sum_{n=1}^{\infty}\Bigl[\frac{\sin 2(n-1)\theta}{4(n-1)}\Bigl(\rho^{2(n-1)}\Bigl(\int_0^\rho t^{-2n+3}{\widehat W}^{(S)}_n\!(t)dt-{\widehat \alpha}_n\Bigr)\\
& -&\rho^{-2(n-1)}\int_0^\rho t^{2n-1}{\widehat W}^{(S)}_n\!(t)dt\Bigr)\Bigr]\qquad {\rm with}\nonumber\\
{\widehat \alpha}_n&=&\int_0^1\left(t^{-2n+3}-t^{2n-1}\right){\widehat W}^{(S)}_n\!(t)dt\qquad {\rm and}\nonumber\\
{\widehat W}^{(S)}_n\!(\rho)&\equiv&\frac{1}{\pi}\int_{-\pi}^{\pi}W^{(S)}(\rho,\theta)\sin\left(2(n-1)\theta\right)d\theta\nonumber
\end{eqnarray}
\noindent It is also easy to convince ourselves that such a procedure also applies for solving the Poisson PDE in case of $n$ independent thermodynamic forces.

\end{document}